\documentclass[onecolumn, floatfix, preprintnumbers, eqsecnum, letterpaper, superscriptaddress, nofootinbib]{revtex4}
\usepackage{graphicx}
\usepackage{microtype}
\usepackage{amsmath}
\usepackage{amssymb}
\usepackage{subfigure}
\usepackage{hyperref}
\usepackage{url}
\usepackage{xcolor}
\usepackage{color}
\usepackage{mathrsfs}
\usepackage{calrsfs}
\usepackage{amsfonts}
\usepackage{eufrak}
\usepackage{tabularx}
\usepackage{eucal}
\usepackage{latexsym}
\usepackage{ragged2e}
\usepackage{epsfig}
\usepackage{textcomp}
\usepackage{inputenc}

\usepackage{caption}
\usepackage{subfigure}

\usepackage{marvosym}
\usepackage{ifsym}

\usepackage{lipsum}
\usepackage{subcaption}

\DeclareCaptionJustification{justified}{\leftskip=0pt \rightskip=0pt \parfillskip=0pt plus 1fil}
\definecolor{vividviolet}{rgb}{0.62, 0.0, 1.0}
\definecolor{amaranth}{rgb}{0.9, 0.17, 0.31}
\definecolor{palatinateblue}{rgb}{0.15, 0.23, 0.89}
\definecolor{brightpink}{rgb}{1.0, 0.0, 0.5}
\definecolor{cornflowerblue}{rgb}{0.39, 0.58, 0.93}
\definecolor{deepcarminepink}{rgb}{0.94, 0.19, 0.22}
\definecolor{radicalred}{rgb}{1.0, 0.21, 0.37}

\hypersetup{ linktoc=all,
	colorlinks, linkcolor={palatinateblue},
	citecolor={brightpink}, urlcolor={amaranth}
}

\graphicspath{{Images/}}

\graphicspath{{Images/}}

\def\@fnsymbol#1{\ensuremath{\ifcase#1\or \ddagger \or  $\textleaf$  \or \dagger
		\else\@ctrerr\fi}}%

\makeatother

\def\sideremark#1{\ifvmode\leavevmode\fi\vadjust{\vbox to0pt{\vss
			\hbox to 0pt{\hskip\hsize\hskip1em
				\vbox{\hsize1.3cm\tiny\raggedright\pretolerance10000
					\noindent #1\hfill}\hss}\vbox to8pt{\vfil}\vss}}}%
\def\beq{\begin{equation}}
	\def\eeq{\end{equation}}

\begin{document}
\title{Investigating shadow of a rotating charged black hole with a cosmological constant immersed in the perfect fluid dark matter }

\author{Zheng-Long Ban}\email{gs.zlban22@gzu.edu.cn}
\address{College of Physics, Guizhou University , Guizhou, China}

\author{Xiao-Jun Gao}\email{gaoxiaojun@gznu.edu.cn}
\address{School of Physics and Electronic Science, Guizhou Normal University, Guiyang 550001, People’s Republic of China College of Physics, Nanjing University of Aeronautics and Astronautics, Nanjing 211106, China}

\author{Jinsong Yang}\email{jsyang@gzu.edu.cn}

\begin{abstract}
In this paper, we mainly investigate the shadow of a rotating charged black hole with a cosmological constant immersed in perfect fluid dark matter. We first obtain the charged spherically symmetric black hole with a cosmological constant solution immersed in perfect fluid dark matter by using the gravitational decoupling method. Based on the mass function seed source of the spherically symmetric solution, we construct a rotating charged black hole with a cosmological constant immersed in perfect fluid dark matter, and study the effects of the perfect fluid dark matter parameter $\alpha$, cosmological constant $\Lambda$, electric charge $Q$, and rotating parameter $a$ on the event horizons of the rotating black hole. We then derive the geodesic equation of photon of the rotating charged black hole with a cosmological constant immersed in perfect fluid dark matter. In addition, we analyze the influences of the four black hole parameters on the effective potential functions of photon, the boundaries, and the deformations of the rotating black hole shadows. Particularly, we find that the effects of the $\alpha$ on the event horizons, the effective potential functions, the boundaries, and the deformations of the black hole shadows are more obvious than the other three parameters ($\Lambda, Q$, and $a$). We expect our results will be useful in the future to relate the theoretical models of perfect fluid dark matter and dark energy with observations of celestial bodies immersed in dark matter and dark energy.

\end{abstract}

\maketitle



\section{Introduction}
General theory of relativity (GR) predicted that black holes (BHs) are formed under gravitational collapse of a sufficiently compact massive object with exerting all its resources to withstand gravity. BHs have been regarded as the most mysterious one of celestial bodies with their remarkable geometric properties. Recently, the astronomical observations of the Laser-Interferometer Gravitational Wave Observatory (LIGO) \cite{LIGOScientific:2016aoc} and the Event Horizon Telescope (EHT) Collaboration \cite{EventHorizonTelescope:2019dse,EventHorizonTelescope:2019ths,EventHorizonTelescope:2019ggy,EHT:2022wkp,EHT:2022wok} have verified the presence of BHs in our Universe. Particularly, the ultra-high angular resolution images of supermassive BHs in the $M87^*$ \cite{EventHorizonTelescope:2019dse,EventHorizonTelescope:2019ths,EventHorizonTelescope:2019ggy} and Sagittarius $A^*$ \cite{EHT:2022wkp,EHT:2022wok} have shown that there is a dark central region surrounded by a bright ring, corresponding to shadow and photon ring of BH. These images may carry some valuable information of the spacetime geometry around BHs. Therefore, the investigations of BH shadows may offer a new window to constrain different gravity models \cite{Mizuno:2018lxz,Stepanian:2021vvk}.

For theoretical analysis, a few pioneering works have been done on the investigation of BH shadows. Synge was the first to demonstrate the shadow of a spherically symmetric BH \cite{Synge:1966okc}. After that, the shadow deformation shape of the Kerr BH was first verified by Bardeen \cite{Bardeen:1973tla}. Vries investigated the apparent shapes of various Kerr-Newman spacetimes \cite{deVries:1999tiy}. Haroon et al. analyzed the influences of perfect fluid dark matter (PFDM) and a cosmological constant on the shadow of a rotating BH \cite{Haroon:2018ryd}. Due to recent astrophysical advances, many authors have comprehensively invested in theoretical investigations of BH shadows \cite{Gao:2024ejs,Meng:2022kjs,Kuang:2024ugn,Meng:2023wgi,Wang:2023vcv,Zhang:2023okw,Yang:2022btw,Zhu:2019ura}. In addition, an interesting review on shadows of a BH was recently done in the literature \cite{Perlick:2021aok}.

Although GR remarkably explains the gravitational interaction at the galactic and cosmological scales, it still lacks to describe two of the major constituents of the Universe--dark matter and dark energy. Our Universe is made up of $27\%$ dark matter, $68\%$ dark energy, and $5\%$ baryonic matter in the standard model of cosmology \cite{Planck:2013pxb}. On the one hand, studies were conducted using dark energy models related to the vacuum energy associated with the cosmological constant ($\Lambda$) \cite{Peebles:2002gy,Koyama:2007rx}. Many research works have demonstrated the significant role of the cosmological constant in a wide range of astrophysical phenomena \cite{Stuchlik:2005euw,Cruz:2004ts,Rezzolla:2003re,Shaymatov:2018azq,Grenon:2009sx}. Subsequently, quintessential scalar fields have also suggested instead of the cosmological constant $\Lambda$ as an alternative form of the dark energy \cite{Wetterich:1987fm,Caldwell:2009zzb}. On the other hand, the most favorable model for dark matter is the cold dark matter model, among primary candidates are Weakly interacting massive particles (WIMPs) \cite{Schumann:2019eaa}. Later, warm and fuzzy dark matter models have also been proposed to fill up the gap of the cold dark matter model at small scales \cite{Tulin:2017ara}.

Above dark matter models belong to the category of PFDM. The perfect fluid model was first investigated in \cite{Kiselev:2002dx,Kiselev:2003ah} and then further work was carried out in \cite{Li:2012zx}. Recently, the PFDM model has been studied using cosmological constant in \cite{Haroon:2018ryd,Xu:2017bpz,Xu:2016ylr}. The shadow and circular geodesics of rotating (charged) BH immersed in PFDM have been analyzed in \cite{Hou:2018avu,Atamurotov:2021hck,Das:2021otl}. However, the \cite{Haroon:2018ryd} only analyzed the effects of PFDM and cosmological constant on a rotating BH shadow. Meanwhile, the \cite{Atamurotov:2021hck} merely investigated the rotating charged BH shadows in the presence of PFDM, while cosmological constant was not taken into account. In the present work, we would like to together consider the effects of PFDM, electric charge, and cosmological constant on a rotating BH shadow. According to the gravitational decoupling method \cite{Ovalle:2017fgl,Mahapatra:2022xea,Ovalle:2021jzf,Contreras:2021yxe,Ovalle:2020kpd,Maurya:2019xcx,Ovalle:2018umz,Heydarzade:2023dof}, the charged spherically symmetric BH with a cosmological constant solution immersed in PFDM is obtained. By utilizing the mass function $\tilde{m}(r)$ of the spherically symmetric solution as the seed source, we derive a new rotating charged BH with a cosmological constant solution in the presence of PFDM.

The paper is organized as follows: in Sect. \ref{section2}, we obtain the rotating charged BH with a cosmological constant solution immersed in PFDM by using the gravitational decoupling method, and analyze the influences of the spin parameter $a$, the cosmological constant $\Lambda$, electric charge $Q$, and PFDM parameter $\alpha$ on the event horizons of rotating BHs, respectively. We investigate the effective potential of photon, boundary, and deformation of shadow in the rotating charged BH with a cosmological constant immersed in PFDM in Sect. \ref{section3}. Finally, Sect. \ref{section4} is for conclusion and discussion. Throughout this paper, we use the geometric units with G = c = 1 unless otherwise specified.
 
\section{The rotating charged BH with a cosmological constant solution immersed in PFDM by the gravitational decoupling method.}
\label{section2}
In this section, by using the gravitational decoupling method \cite{Ovalle:2017fgl}, we first derive the charged spherically symmetric black hole with a cosmological constant solution in the presence of PFDM. On the basis of the mass function of the spherically symmetric seed solution, we further obtain a new rotating charged BH with a cosmological constant solution immersed in PFDM. Finally, we numerically analyze the influences of the corresponding BH parameters on the event horizon of the new rotating BH.

\subsection{Static spherically symmetric charged black hole solution with a cosmological constant immersed in PFDM}

Based on the gravitational decoupling method \cite{Ovalle:2017fgl}, we follow the calculational process of the reference \cite{Ovalle:2021jzf} to generate the spherically symmetric charged BH with a cosmological constant solution immersed in PFDM. We start from the standard Einstein field equation with a cosmological constant
\begin{equation}
R_{\mu \nu}-\frac{1}{2} R g_{\mu \nu}+\Lambda g_{\mu\nu}=8 \pi T_{\mu \nu}.\label{fieldequation}
\end{equation}
Taking into account the static and spherically symmetric spacetime, the line element can be written in the following form \cite{Visser1995}: 
\begin{equation}
ds^{2}=-e^{\Phi(r)}\left(1-\frac{2  m(r)}{r}\right) d t^{2}+\frac{d r^{2}}{1-\frac{2  m(r)}{r}}+r^2(d\theta^2+\sin^2\theta d\phi^2),\label{lineelement}
\end{equation}
where $\Phi(r)$ denotes a metric function,and $m(r)$ represents the Misner Sharp mass function, which measures the amount of energy within a sphere of areal radius $r$.
Generally, one can set the metric function $\Phi(r)=0$. By substituting Eq.  (\ref{lineelement}) into  Eq.(\ref{fieldequation}), the field equations are given by
\begin{align}
-\dfrac{2m'(r)}{r^2}+\Lambda=8\pi T_0~^0,\label{firstEq}\\
-\dfrac{2m'(r)}{r^2}+\Lambda=8\pi T_1~^1,\label{secondEq}\\
-\dfrac{m''(r)}{r}+\Lambda=8\pi T_2~^2,\label{thirdEq}
\end{align}
where the prime symbol represents the derivative with respect to $r$. From Eqs. (\ref{firstEq})-(\ref{thirdEq}), one not only discovers that the equation of state satisfies the relation $T_0~^0=-T_1~^1$, but also that the slope of the mass function $m(r)$ is a linear function \cite{Ovalle:2021jzf}. Based on this latter characteristic feature, any solution $m(r)$ of the system  Eqs. (\ref{firstEq})-(\ref{thirdEq}) can be coupled with a second one $m_s(r)$ to generate a new solution $\tilde m(r)$ as
 \begin{align}
 m(r)\longrightarrow \tilde{m}(r)=m(r)+m_s(r).\label{newtildemr}
 \end{align}
 
We consider dark matter that is minimally coupled to gravity to generate a charged BH with a cosmological constant solution immersed in dark matter. Hence, the new energy-momentum tensor $\tilde T_{\mu\nu}$ is assumed as
\begin{align}
\tilde T_{\mu\nu}=T^M_{\mu\nu}-T^{DM}_{\mu\nu},\label{MDM}
\end{align}
where $T^M_{\mu\nu}$ denotes the energy-momentum tensor of the electromagnetic field as follows \cite{Heydarzade:2017wxu}: 
\begin{equation}
(T^\mu{}_ \nu)^{M}=\operatorname{diag}\left(-\frac{Q^2}{8\pi r^4},~-\frac{Q^2}{8\pi r^4},~\frac{Q^2}{8\pi r^4},~\frac{Q^2}{8\pi r^4}\right), \label{matterT}
\end{equation}
with $Q$ is the electric charge; while $T_{\mu \nu}^{DM}$ is regarded as the energy-momentum tensor of the PFDM, which is given by \cite{Li:2012zx,Atamurotov:2021hck} 
\begin{equation}
\left(T^{\mu}{ }_{\nu}\right)^{D M}=\operatorname{diag}\left(-\rho, P_{r}, P_{\theta}, P_{\phi}\right);~~ \rho=-P_r;~~ P_{\theta}=P_{\phi},\label{DMT}
\end{equation}
where $\rho$ and $P_i$ correspond to the density and pressure, respectively.

According to Eqs. (\ref{firstEq})-(\ref{DMT}), the different
components of field equations take the form
\begin{align}
-\dfrac{2\tilde{m}'(r)}{r^2}+\Lambda&=8\pi\left(-\dfrac{Q^2}{8\pi r^4}+\rho\right),\label{newfirstEq}\\
-\dfrac{2\tilde{m}'(r)}{r^2}+\Lambda&=8\pi\left(-\dfrac{Q^2}{8\pi r^4}-P_r\right),\label{newsecondEq}\\
-\dfrac{2\tilde{m}''(r)}{r}+\Lambda&=8\pi\left(\dfrac{Q^2}{8\pi r^4}-P_{\theta}\right).\label{newthirdEq}
\end{align}
Following the literature \cite{Li:2012zx,Atamurotov:2021hck}, taking the equation of state for the PFDM as $P_{\theta}=P_{\phi}=\rho(\delta-1)$ with $\delta$ being a constant. From Eq. (\ref{newfirstEq}) and Eq. (\ref{newthirdEq}), we get
\begin{align}
\dfrac{\tilde{m}''(r)}{r^2}-\Lambda+\dfrac{Q^2}{r^4}=(\delta-1)\left(-\dfrac{2\tilde{m}'(r)}{r^2}+\Lambda+\dfrac{Q^2}{r^4}\right).\label{newfieldEq}
\end{align}
For different values of $\delta$, the equation can have different solutions \cite{Li:2012zx}. We merely focus on the solution of particular interest when $\delta=3/2$. The Eq. (\ref{newfieldEq}) is simplified to
\begin{align}
r\tilde{m}''(r) +\tilde{m}'(r)+\dfrac{1}{2}\dfrac{Q^2}{r^2}-\dfrac{3}{2}\Lambda r^3=0.\label{simplifiedEq}   
\end{align}
The solution to the above equation is obtained as
\begin{align}
\tilde{m}(r)=c_2-\dfrac{Q^2}{2r}+\dfrac{\Lambda}{6}r^3+c_1\ln{r}.\label{solutionEq}    
\end{align}
where $c_i$ are integration constants. We set $c_1=-\alpha/2$ and $c_2=M+(\alpha\ln{|\alpha|})/2$ to generate a charged BH with a cosmological constant solution immersed in PFDM, and by substituting Eq. (\ref{solutionEq}) into Eq. (\ref{lineelement}), we can obtain
\begin{align}
f(r)=1-\dfrac{2\tilde{m}(r)}{r}=1-\dfrac{2M}{r}+\dfrac{Q^2}{r^2}-\dfrac{\Lambda}{3}r^2+\dfrac{\alpha}{r}\ln{\dfrac{r}{|\alpha|}},\label{QDMmetric}  
\end{align}
where $M$ is the BH mass parameter, and $\alpha$ denotes the PFDM parameter. The solution in Eq. (\ref{QDMmetric}) is consistent with the results from the reference \cite{Atamurotov:2021hck} when $\Lambda=0$.

\subsection{The rotating charged BH with a cosmological constant solution immersed in PFDM}
In order to generate the rotating version of a charged BH with a cosmological constant solution immersed in PFDM in Eq.(\ref{QDMmetric}), we follow the strategy described in Ref. \cite{Contreras:2021yxe}. In Boyer-Lindquist coordinates, the Gurses-Gursey metric is given by \cite{Gurses:1975vu}

 \begin{equation}
d s^{2}= {\left[1-\frac{2 r \tilde m(r)}{\rho^{2}}\right] d t^{2}+\frac{4 a r \tilde m(r) \sin ^{2} \theta}{\rho^{2}} d t d \phi } \\
 -\frac{\rho^{2}}{\Delta} d r^{2}-\rho^{2} d \theta^{2}-\frac{\Xi  \sin ^{2} \theta}{\rho^{2}} d \phi^{2},\label{ds28}
\end{equation}
where
\begin{align}
\rho^{2}  =r^{2}+a^{2} \cos ^{2} \theta,~\Delta  =r^{2}-2 r \tilde m(r)+a^{2} ,~\Xi   =\left(r^{2}+a^{2}\right)^{2}-\Delta a^{2} \sin ^{2} \theta,  
\end{align}
and $a$ is the angular velocity of rotation. Next, the spherically symmetric mass function $\tilde{m}(r)$ in Eq.(\ref{solutionEq}) is regarded as a seed source to generate the new rotating BH solution. Plugging Eq. (\ref{solutionEq}) into the metric (\ref{ds28}) leads to

\begin{align}
	d s_{new}^{2} &= -\frac{\Delta_{r} - a^{2} \sin^{2} \theta}{\Sigma} d t^{2} 
	+ \frac{\Sigma}{\Delta_{r}} d r^{2} 
	+ \Sigma d \theta^{2}
	+ \frac{\sin^{2} \theta}{\Sigma} \left( \left(r^{2} + a^{2}\right)^{2} - \Delta_{r} a^{2} \sin^{2} \theta \right) d \phi^{2} \notag \\
	&\quad - \frac{2 a \sin^{2} \theta}{\Sigma} \left( \left(r^{2} + a^{2}\right) - \Delta_{r} \right) d t d \phi \label{knd8}
\end{align}

where
\begin{align}
\Delta_{r}=a^{2}+Q^{2}-2 M r+r^{2}-\frac{r^{4} \Lambda}{3}+r \alpha\ln \left(\frac{r}{|\alpha|}\right),~\Sigma=r^{2}+a^{2} \cos ^{2} \theta.\label{deltafunction}   
\end{align}
{Eq. (\ref{knd8}) can reduce to the result of the Ref. \cite{Atamurotov:2021hck} as $\Lambda=0$. It is worth mentioning that Eq. (\ref{knd8}) can't reduce to the Kerr-de Sitter solution for $Q=\alpha=0$. The corresponding reason analysis has been demonstrated in Ref. \cite{Ovalle:2021jzf}. Therefore, we first derive the rotating charged BH with a cosmological constant solution Eq.(\ref{knd8}) immersed in PFDM by the gravitational decoupling.

\begin{figure}[htbp]
\begin{minipage}[t]{0.48\linewidth}
\centering
\includegraphics[width=3.2in]{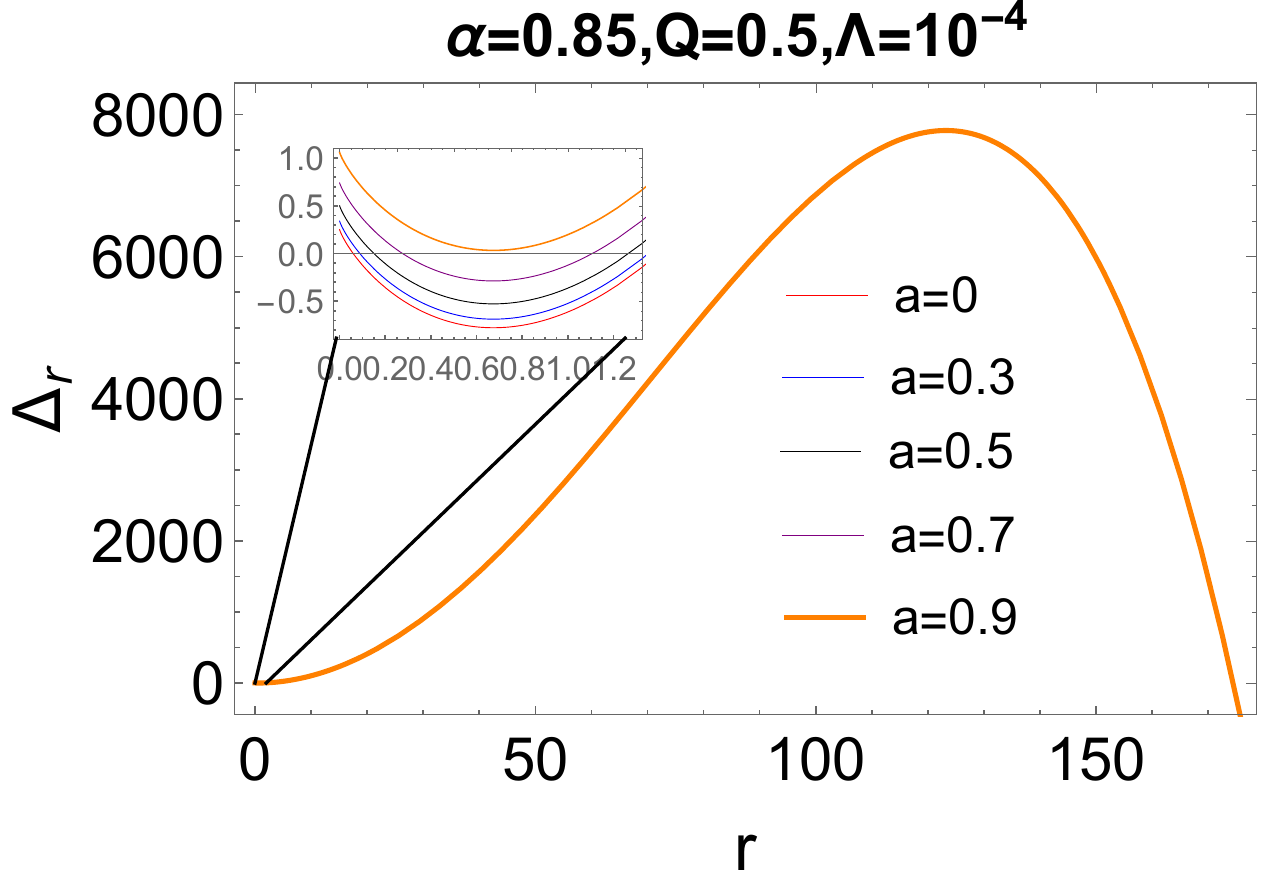}
\caption{The event horizons as a function of the $r$ for the different $a$ values, setting $\alpha=0.85, Q=0.5, \Lambda=10^{-4}$ and $M=1$.}
\label{fig1}
\end{minipage}
\hspace{0.1cm}
\begin{minipage}[t]{0.48\linewidth}
\centering
\includegraphics[width=3.2in]{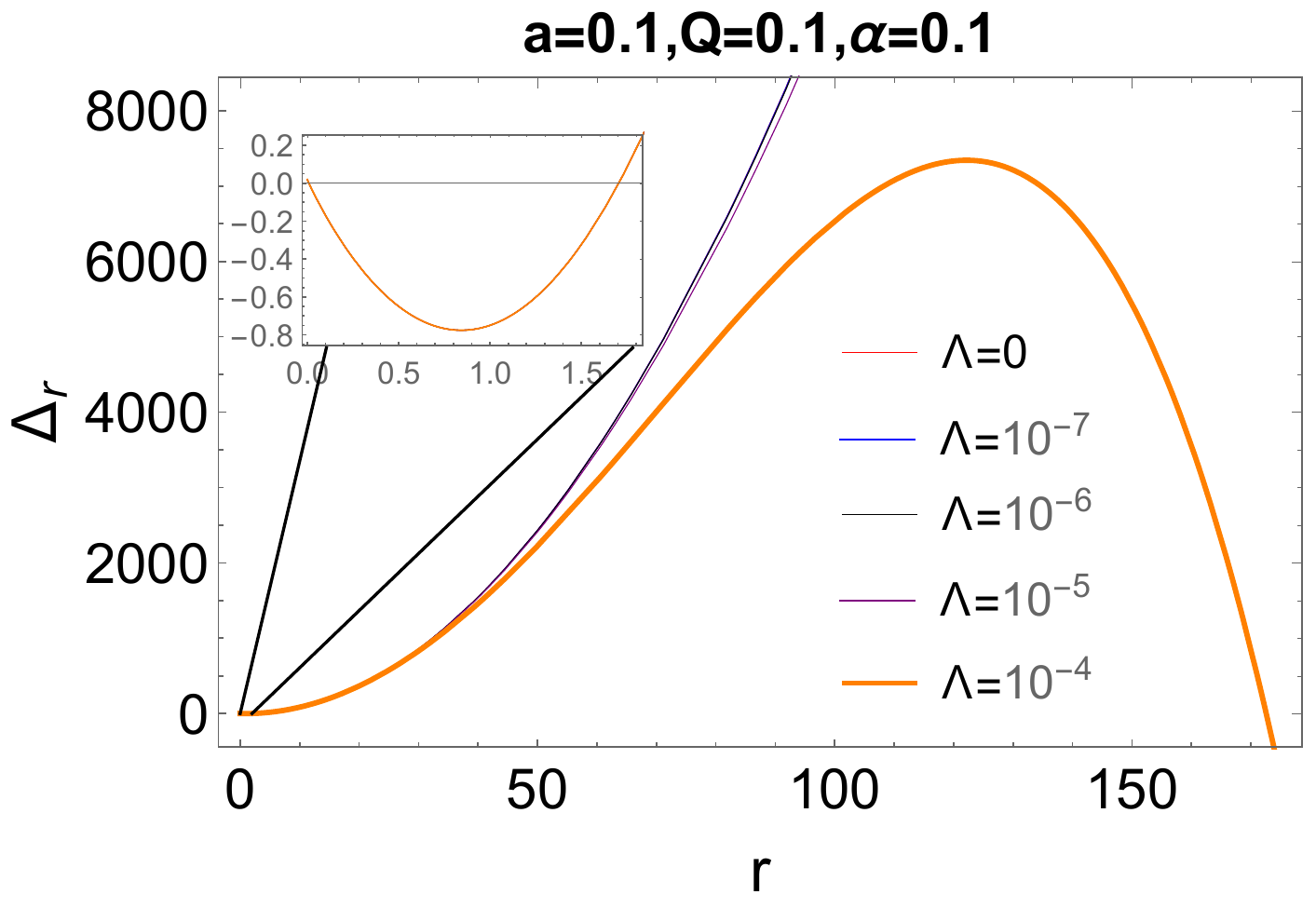}\\
\caption{The event horizons as a function of the $r$ for the different $\Lambda$ values, setting $a=0.5, Q=0.1, \alpha=0.1$ and $M=1$.}
\label{fig2}
\end{minipage}\\
\begin{minipage}[t]{0.48\linewidth}
\centering
\includegraphics[width=3.2in]{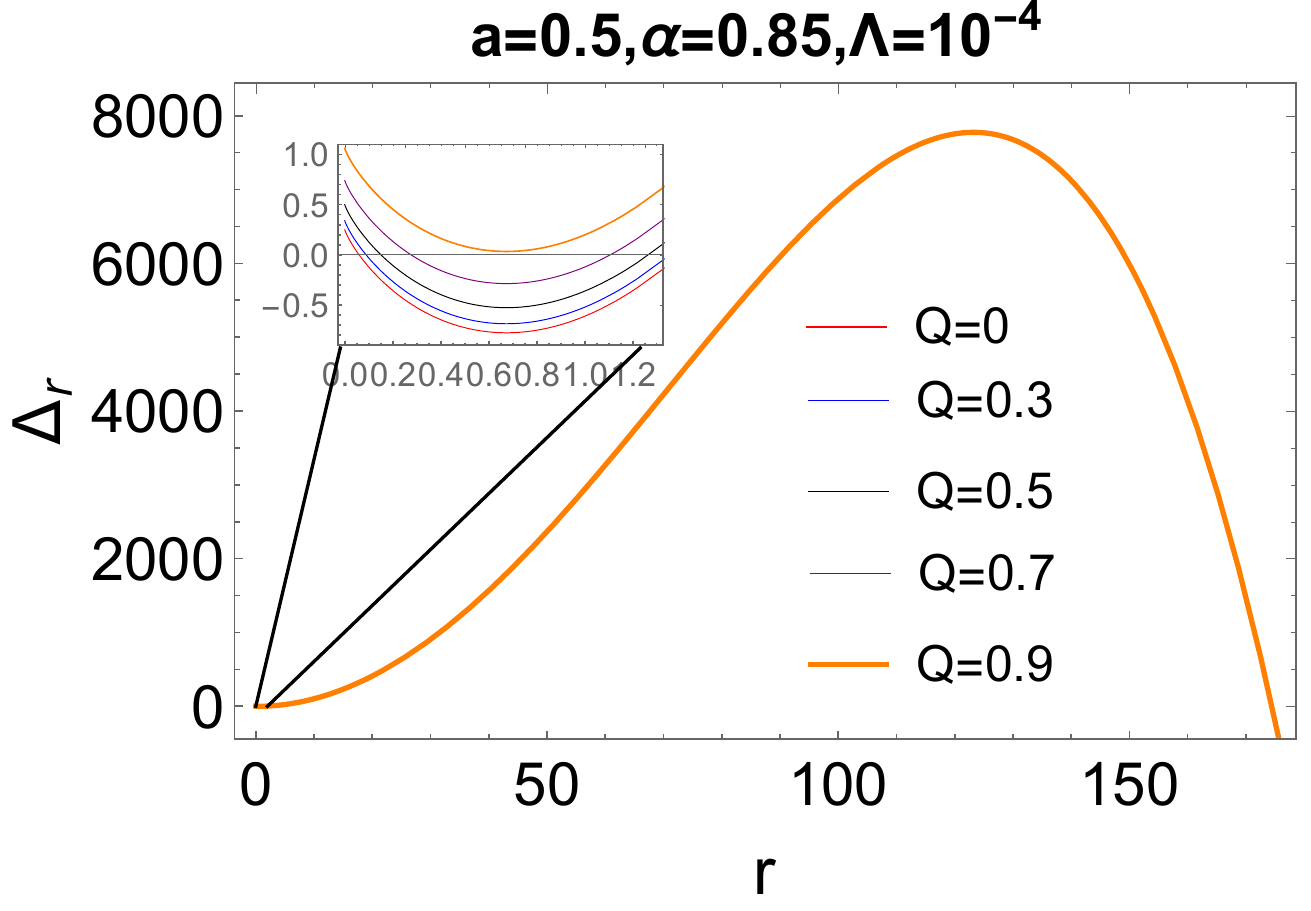}
\caption{The event horizons as a function of the $r$ for the different $Q$ values, setting $a=0.5, \alpha=0.85, \Lambda=10^{-4}$ and $M=1$.}
\label{fig3}
\end{minipage}
\hspace{0.1cm}
\begin{minipage}[t]{0.48\linewidth}
\centering
\includegraphics[width=3.2in]{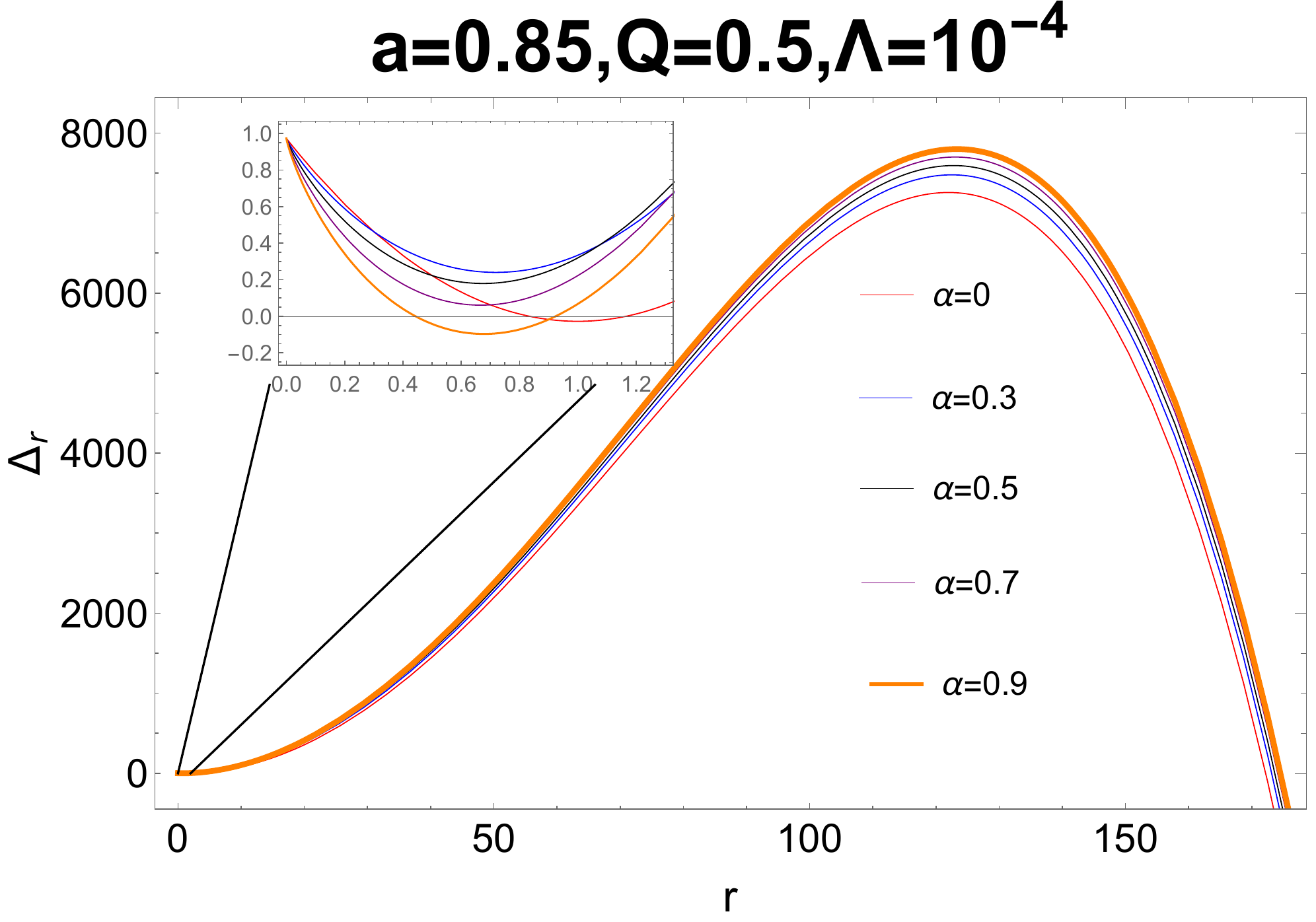}\\
\caption{The event horizons as a function of the $r$ for the different $\alpha$ values, setting $a=0.85, Q=0.5, \Lambda=10^{-4}$ and $M=1$.}
\label{fig4}
\end{minipage}
\end{figure}
In general, the event horizon is determined by solving $\Delta_{r}=0$ for a rotating BH. The $\Delta_{r}=0$ in Eq.(\ref{deltafunction}) doesn't yield an analytical solution due to the existence of the logarithmic term $\ln r$. Hence, we can't analyze the event horizon of the new rotating BH by the analytical expression. However, we are able to resort to numerical methods. For convenience in analysis, we set $M=1$. We have plotted the $\Delta_{r}$ as the functions of $r$ for the different values of BH parameters ($a, \Lambda, Q$, and $\alpha$) in the Figs. \ref{fig1}-\ref{fig4}. From Figs. \ref{fig1} and \ref{fig3}, it is clearly shown that the event horizons decrease with the increase of $a$ or $Q$. The $\Lambda$ hardly affects the event horizons in Fig. \ref{fig2}. The influences of $\alpha$ on the event horizons are nonlinear in Fig. \ref{fig4}. In addition, we find that the effects of $\alpha$ on the cosmological event horizons are more obvious than the other three parameters ($a, Q$, and $\Lambda$).
\section{The shadow by the rotating charged BH with a cosmological constant immersed in PFDM }
\label{section3}
In this section, in order to study the shadow cast by the rotating charged BH with a cosmological constant immersed in PFDM, we first derive the geodesic equation of a photon, and then investigate the influences of the BH parameters ($\alpha, \Lambda, Q$, and $a$) on the boundaries and the deformations of the rotating BH shadows. Meanwhile, we also analyze the effects of the four BH parameters on the effective potential functions of a photon.
\subsection{Geodesic equation of photon}
The Hamilton-Jacobi equation is written as \cite{Meng:2022kjs,Carter:1968rr}
\begin{equation}
\frac{\partial S}{\partial \lambda }=-\frac{1}{2} g^{\mu \nu} \frac{\partial S}{\partial x^{\mu}} \frac{\partial S}{\partial x^{\nu}},\label{s31}
\end{equation}
where $\lambda$ represents the affine parameter of the geodesic, and $S$ denotes the Jacobi action. With the two conserved quantities, the energy $E$ and angular momentum $L$ of the particle,and $S$ can be expressed in a separable form as follows:
\begin{equation}
S=\frac{1}{2} m^{2} \lambda -E t+L \phi+S_{r}(r)+S_{\theta}(\theta),  \label{s32}            
\end{equation}
where $m$ is the mass of the test particle, which is assumed to be zero for a null geodesic and one for a timelike geodesic. We mainly focus on photons ($m=0$) in this paper. The quantities $E$ and $L$ are given by
\begin{align}
E&=-g_{a b}\left(\frac{\partial}{\partial t}\right)^{a}\left(\frac{\partial}{\partial \lambda }\right)^{b}=-g_{00} \dot{t}-g_{03} \dot{\phi}, \label{likenull13}\\
L&=g_{a b}\left(\frac{\partial}{\partial \phi}\right)^{a}\left(\frac{\partial}{\partial\lambda }\right)^{b}=g_{33} \dot{\phi}+g_{03} \dot{t} ,\label{likenull14}
\end{align}
respectively. By solving the Eqs. (\ref{likenull13}) and (\ref{likenull14}), we obtain
\begin{align}
\Sigma \frac{d t}{d \lambda }&=\frac{r^{2}+a^{2}}{\Delta_{r}}\left(E\left(r^{2}+a^{2}\right)-a L\right)-a\left(a E \sin ^{2} \theta-L\right),\label{likenull1}\\ 
\Sigma \frac{d \phi}{d \lambda }&=\frac{a}{\Delta_{r}}\left(E\left(r^{2}+a^{2}\right)-a L\right)-\left(a E-L \csc ^{2} \theta\right),\label{likenull2}
\end{align}
where $\Delta_{r}$ and $\Sigma$ have been given in Eq. (\ref{deltafunction}). Utilizing Eqs. (\ref{likenull1}) and (\ref{likenull2}), then substituting Eq. (\ref{s32}) into Eq. (\ref{s31}) and performing variable separation, the components for radial and polar motion are given by
\begin{align}
 \left(\Sigma \frac{d r}{d \lambda }\right)^{2}=\left(E\left(r^{2}+a^{2}\right)-a L\right)^{2}-\mathcal{K} \Delta_{r}&\equiv R(r), \label{likenull3}\\
\left(\Sigma \frac{d \theta}{d \lambda }\right)^{2}=-\csc ^{2} \theta\left(a E \sin ^{2} \theta-L\right)^{2}+\mathcal{K}
&\equiv\Theta(\theta) , \label{likenull4} 
\end{align}
where $\mathcal{K}$ is the Carter constant. To further simplify, we set $\xi=L / E, \eta=\mathcal{K} / E^{2}$. Eqs. (\ref{likenull3}) and (\ref{likenull4}) are simplified to
\begin{align}
 R(r)=\left(\left(r^{2}+a^{2}\right)-a \xi\right)^{2}-\eta \Delta_{r}, \label{likenull5}\\
\Theta(\theta)=-\csc ^{2} \theta\left(a \sin ^{2} \theta-\xi\right)^{2}+\eta .\label{likenull6} 
\end{align} 

The photon sphere orbit satisfies :
\begin{equation}
R\left(r_{s}\right)=0, \left.\quad \frac{d R(r)}{d r}\right|_{r=r_{s}}=0 ,\label{likenull7}
\end{equation}

\begin{equation}
	\begin{split}
		\left.\frac{d^{2} R(r)}{d r^{2}}\right|_{r=r_{s}}  >0 ,
	\end{split}\label{Rr}
\end{equation}

where $r_s$ is the photon sphere radius. Putting Eq. (\ref{likenull5}) into Eq. (\ref{likenull7}), $\xi$ and $\eta$ are obtained
\begin{align}
a\xi&=\left(r_{s}^{2}+a^{2}\right)-\frac{4 r_{s} \Delta_{r}\left(r_{s}\right)}{\Delta_{r}^{\prime}\left(r_{s}\right)},\label{likenull8}\\
\eta&=\frac{16 r_{s}^{2} \Delta_{r}\left(r_{s}\right)}{\left(\Delta_{r}^{\prime}\left(r_{s}\right)\right)^{2}},\label{likenull9} 
\end{align}
respectively. Assuming $\Theta(\theta) \ge 0$ for $\theta$ within the interval $[0, \pi]$, and using Eq. (\ref{likenull8}) and Eq. (\ref{likenull9}), we can derive the inequality characterizing the region $\mathcal{C}$ related to the photon sphere
\begin{equation}
\mathcal{C}:\left(4 r_{s} \Delta_{r}\left(r_{s}\right)-\Sigma \Delta_{r}^{\prime}\left(r_{s}\right)\right)^{2} \le 16 a^{2} r_{s}^{2} \Delta_{r}\left(r_{s}\right) \Delta _{\theta} \sin ^{2} \theta.
\end{equation} 
which reduces to the static BH as $a=0$, and the following condition is satisfied
\begin{equation}
4 r_{s} \Delta_{r}\left(r_{s}\right)=r_{s}^{2} \Delta_{r}^{\prime}\left(r_{s}\right).\label{rsdeltaEq}
\end{equation} 
The Eq. (\ref{rsdeltaEq}) can yield $r_s = 3M$ when $a = Q = \alpha = \Lambda = 0$.

\begin{figure}[htbp]
\begin{minipage}[t]{0.48\linewidth}
\centering
\includegraphics[width=3in]{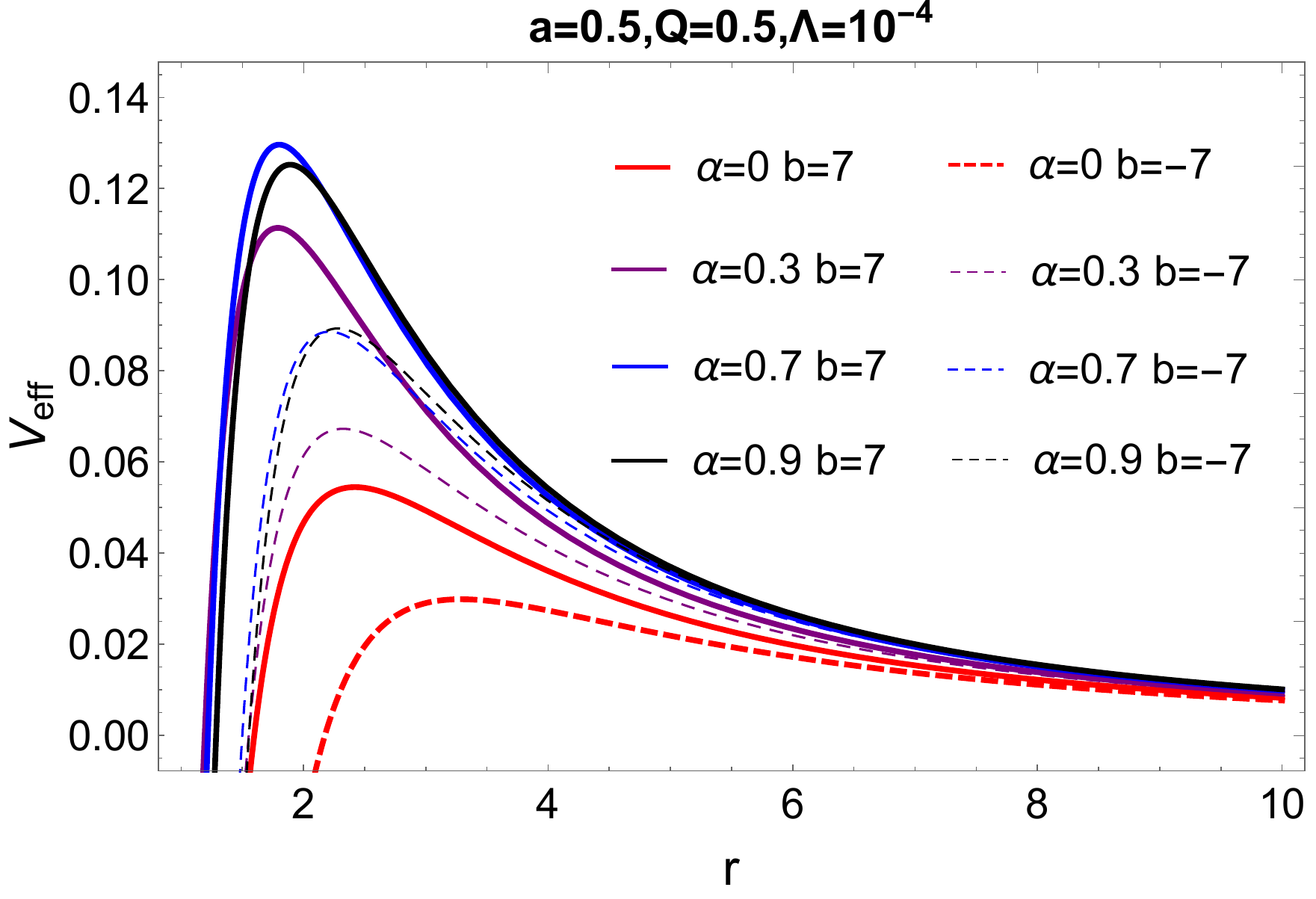}
\caption{The effective potential $V_{eff}(r)$ as a function of the $r$ for the different $\alpha$ values, setting $a=0.5, Q=0.5, \Lambda=10^{-4}$ and $M=1$.}
\label{fig5}
\end{minipage}
\hspace{0.1cm}
\begin{minipage}[t]{0.48\linewidth}
\centering
\includegraphics[width=3in]{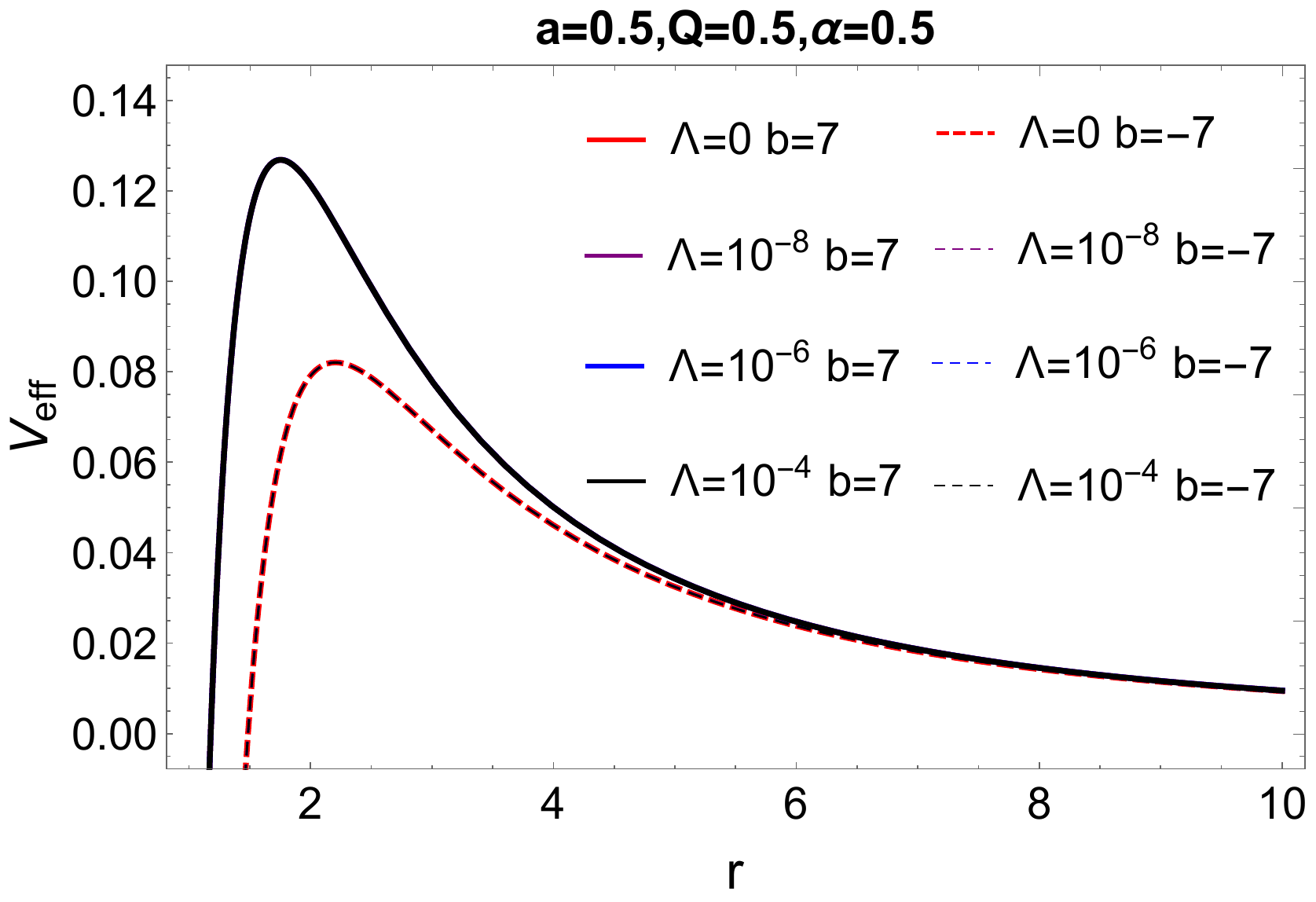}\\
\caption{The effective potential $V_{eff}(r)$ as a function of the $r$ for the different $\Lambda$ values, setting $a=0.5, Q=0.5, \alpha=0.5$ and $M=1$.}
\label{fig6}
\end{minipage}\\
\begin{minipage}[t]{0.48\linewidth}
\centering
\includegraphics[width=3in]{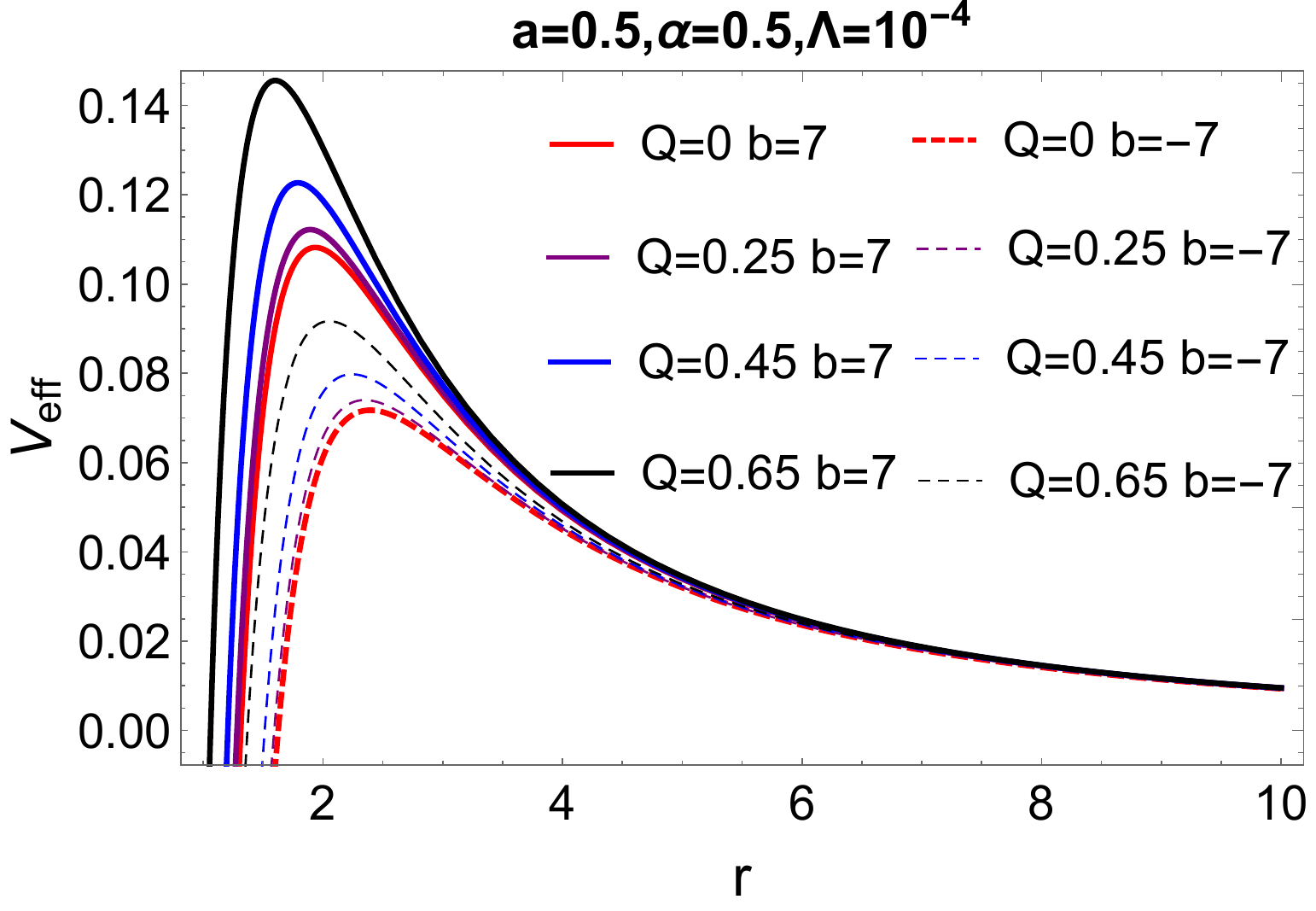}
\caption{The effective potential $V_{eff}(r)$ as a function of the $r$ for the different $Q$ values, setting $a=0.5, \alpha=0.5, \Lambda=10^{-4}$ and $M=1$.}
\label{fig7}
\end{minipage}
\hspace{0.1cm}
\begin{minipage}[t]{0.48\linewidth}
\centering
\includegraphics[width=3in]{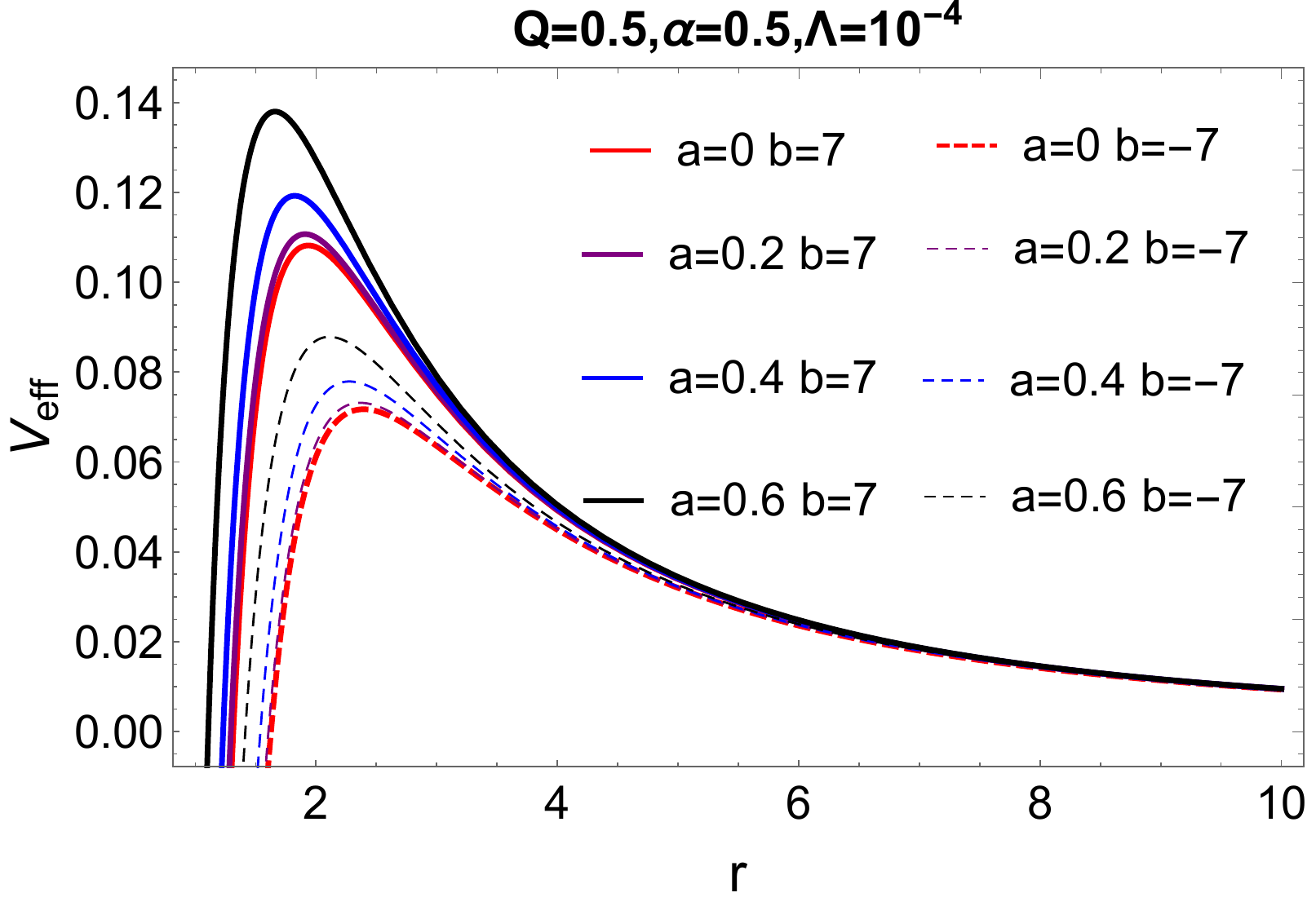}\\
\caption{The effective potential $V_{eff}(r)$ as a function of the $r$ for the different $a$ values, setting $Q=0.5, \alpha=0.5, \Lambda=10^{-4}$ and $M=1$.}
\label{fig8}
\end{minipage}
\end{figure}
\subsection{Effective Potential of Photon}

Without loss of generality, we consider the light traveling on the equatorial plane of this compact object, i.e., $\theta=\pi/2$ and $\dot{\theta}=0$, and then $\mathcal{K}$ vanishes in Eq. (\ref{likenull4}). From Eq. (\ref{likenull3}), the effective potential function is defined by \cite{Hsiao:2019ohy}.
\begin{align}
\frac{1}{b^{2}}=\frac{\dot{r}^{2}}{L^{2}}+V_{eff}(r),\label{effectivefun}    
\end{align}
and 
\begin{align}
V_{eff}(r)=\frac{1}{r^{2}}\left[1-\frac{a^{2}}{b^{2}}+\left(-\frac{2 M}{r}+\frac{Q^{2}}{r^{2}}+\frac{\alpha\ln \left(\frac{r}{|\alpha|}\right)\text { } }{r}-\frac{r^{2} \Lambda}{3}\right)\left(1-\frac{a}{b_s}\right)^{2}\right].\label{veff}    
\end{align}
where $b\equiv E/L$ is a constant called the impact parameter, and $b_s=\pm b$. The $b_s>0$ is referred to as direct orbits, while $b_s<0$ is regarded as retrograde orbits. 
We choose $b_s=\pm 7$ to respectively investigate the influence of the four BH parameters ($\alpha, \Lambda, Q,$ and $a$) on the effective potential function Eq. (\ref{veff}) in Figs. \ref{fig5}-\ref{fig8}. From Fig. \ref{fig5}, the peak values of effective potential first increase as $\alpha$ increases, and then decrease. In Figs. \ref{fig7} and \ref{fig8}, we find that a larger $Q$ or $a$ leads to a larger peak effective potential at the smaller radius in the direct or retrograde orbits. The cosmological constant $\Lambda$ doesn't affect the effective potential function in Fig. \ref{fig6}. The effects of the PFDM parameters $\alpha$ on peak effective potential are more significant in comparison with $Q$, $a$ and $\Lambda$.

\subsection{The BH shadow observed at a finite distance}

For the rotating BH solution Eq. (\ref{knd8}), the corresponding spacetime is not asymptotically flat due to the presence of the cosmological constant term. If we want to investigate the shadow of the rotating BH Eq. (\ref{knd8}), the distance from the light source and the observer to the BH needs to be finite. In these realistic universe circumstances, the distance between celestial bodies is finite. For the light rays that issue from the position of an observer into the past, the initial direction is determined by two angles in the observer’s sky: a colatitude angle and an azimuthal angle. Then, we consider an observer at position $(r_O, \theta_O)$ in Boyer-Lindquist coordinates. To determine the boundary of the shadow, we choose the orthonormal tetrad \cite{Meng:2022kjs,Carter:1968rr}.

\begin{align}
e_{0}&=\left.\frac{\left(r^{2}+a^{2}\right) \partial_{t}+a \partial_{\phi}}{\sqrt{\Delta_{r} \Sigma}}\right|_{\left(r_{o}, \theta_{o}\right)},~~e_{1}=\left.\frac{1}{\sqrt{\Sigma}} \partial_{\theta}\right|_{\left(r_{o}, \theta_{o}\right)}, \label{tetrade01}\\
e_{2}&=-\left.\frac{\partial_{\phi}+a \sin ^{2} \theta \partial_{t}}{\sqrt{\Sigma} \sin \theta}\right|_{\left(r_{o}, \theta_{o}\right)},~~e_{3}=-\left.\sqrt{\frac{\Delta_{r}}{\Sigma}} \partial_{r}\right|_{\left(r_{o}, \theta_{o}\right)}\label{tetrade23}
\end{align}
at the observation event in the domain of outer communication. Here, $e_0$ represents the observer’s 4-velocity, and $e_3$ corresponds to the spatial direction pointing towards the BH. $e_0\pm e_3$ are tangent to the principal null directions. The photon trajectory in the coordinate system is described as $\lambda(s)=(t(s), r(s), \theta(s), \phi(s))$, and the tangent vector to the photon trajectory is given by

\begin{equation}
\dot{\lambda}=\dot{t} \partial_{t}+\dot{r} \partial_{r}+\dot{\theta} \partial_{\theta}+\dot{\phi} \partial_{\phi} ,\label{a322}
\end{equation}
Using the celestial coordinates $\alpha$ and $ \beta$, the tangent vector at the observation event can be written as 
\begin{equation}
\dot{\lambda}=\gamma\left(-e_{0}+\sin \alpha \cos \beta e_{1}+\sin \alpha \sin \beta e_{2}+\cos \alpha e_{3}\right),\label{a323}
\end{equation}
where the $\gamma$ is given by
\begin{equation}
\gamma=g\left(\dot{\lambda}, e_{0}\right)=\left.\frac{a L-E\left(r^{2}+a^{2}\right)}{\sqrt{\Delta_{r} \Sigma}}\right|_{\left(r_{o}, \theta_{o}\right)} . 
\end{equation}
By comparing Eq. (\ref{a322}) and Eq. (\ref{a323}), one can obtain:
\begin{align}
\sin \alpha&=\left.\sqrt{1-\left[\frac{\dot{r} \Sigma}{E\left(r^{2}+a^{2}\right)-a L}\right]^{2}}\right|_{\left(r_{o}, \theta_{o}\right)}, \label{sinalpha}\\
\sin \beta&=\left.\frac{\sin \theta}{\sqrt{\Delta_{r}} \sin \alpha}\left[\frac{\dot{\phi} \Sigma \Delta_{r}}{E\left(r^{2}+a^{2}\right)-a L}-a\right]\right|_{\left(r_{o}, \theta_{o}\right)}.\label{sinbeta}
\end{align}
Utilizing Eq. (\ref{likenull3}) and Eq. (\ref{likenull2}), Eq. (\ref{sinalpha}) and Eq. (\ref{sinbeta}) are rewritten as
\begin{align}
\sin \alpha&=\left.\frac{\sqrt{\eta\left(r_{s}\right) \Delta_{r}}}{r^{2}+a^{2}-a \xi\left(r_{s}\right)}\right|_{\left(r_{o}, \theta_{o}\right)},\label{sinalphanew}\\
\sin\beta&=\left.\frac{\sqrt{\Delta_{r}} \sin \theta}{\sin \alpha}\left[\frac{a-\xi\left(r_{s}\right) \csc ^{2} \theta}{a \xi\left(r_{s}\right)-\left(r^{2}+a^{2}\right)}\right]\right|_{\left(r_{o}, \theta_{o}\right)}.\label{sinbetanew}
\end{align}
\begin{figure}[htbp]
\begin{minipage}[t]{0.48\linewidth}
\centering
\includegraphics[width=3in]{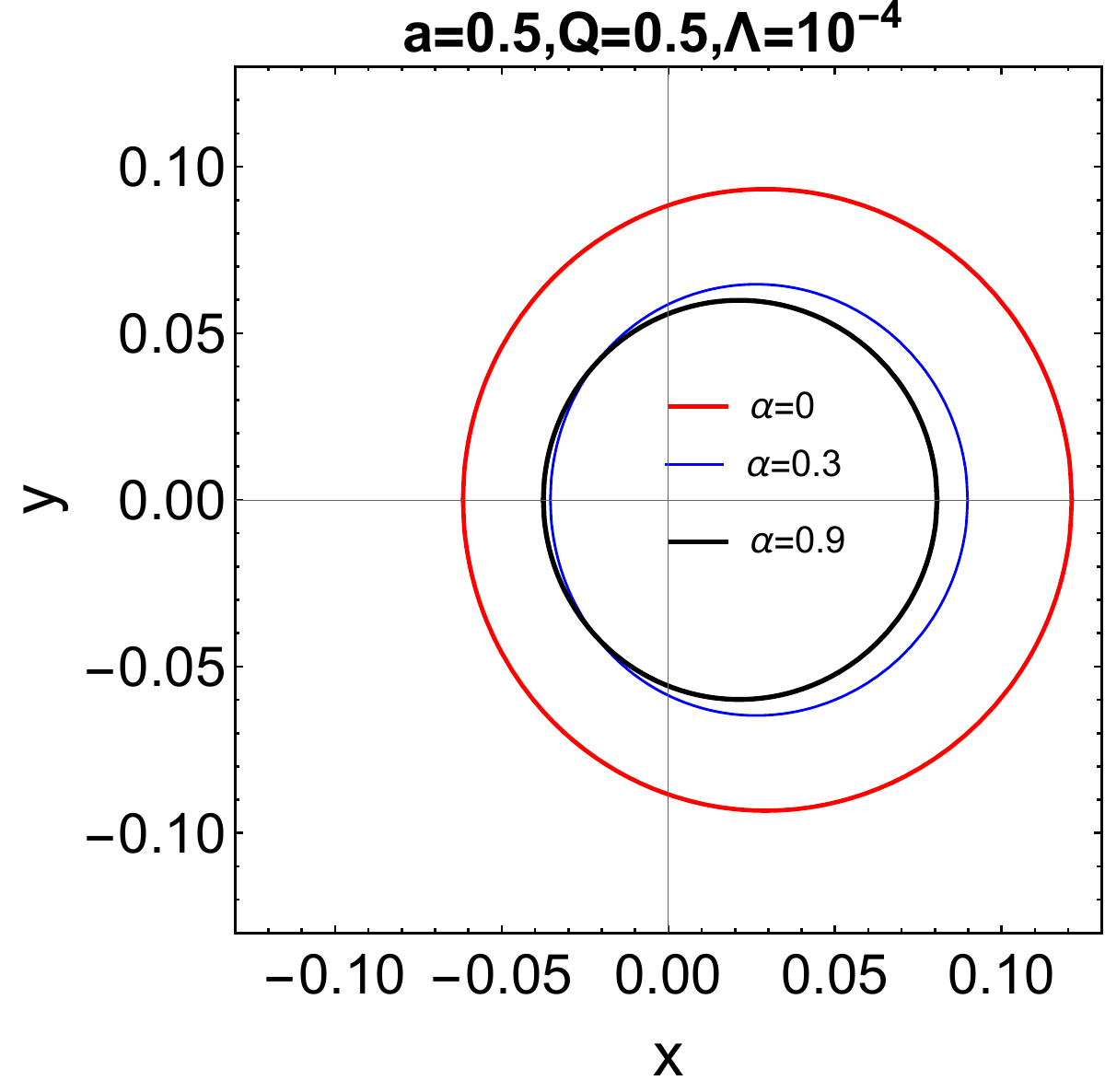}
\caption{Boundary of the black hole shadow for different values of the $\alpha$, setting $a=0.5, Q=0.5, \Lambda=10^{-4}$ and $M=1$.}
\label{fig9}
\end{minipage}
\hspace{0.1cm}
\begin{minipage}[t]{0.48\linewidth}
\centering
\includegraphics[width=3in]{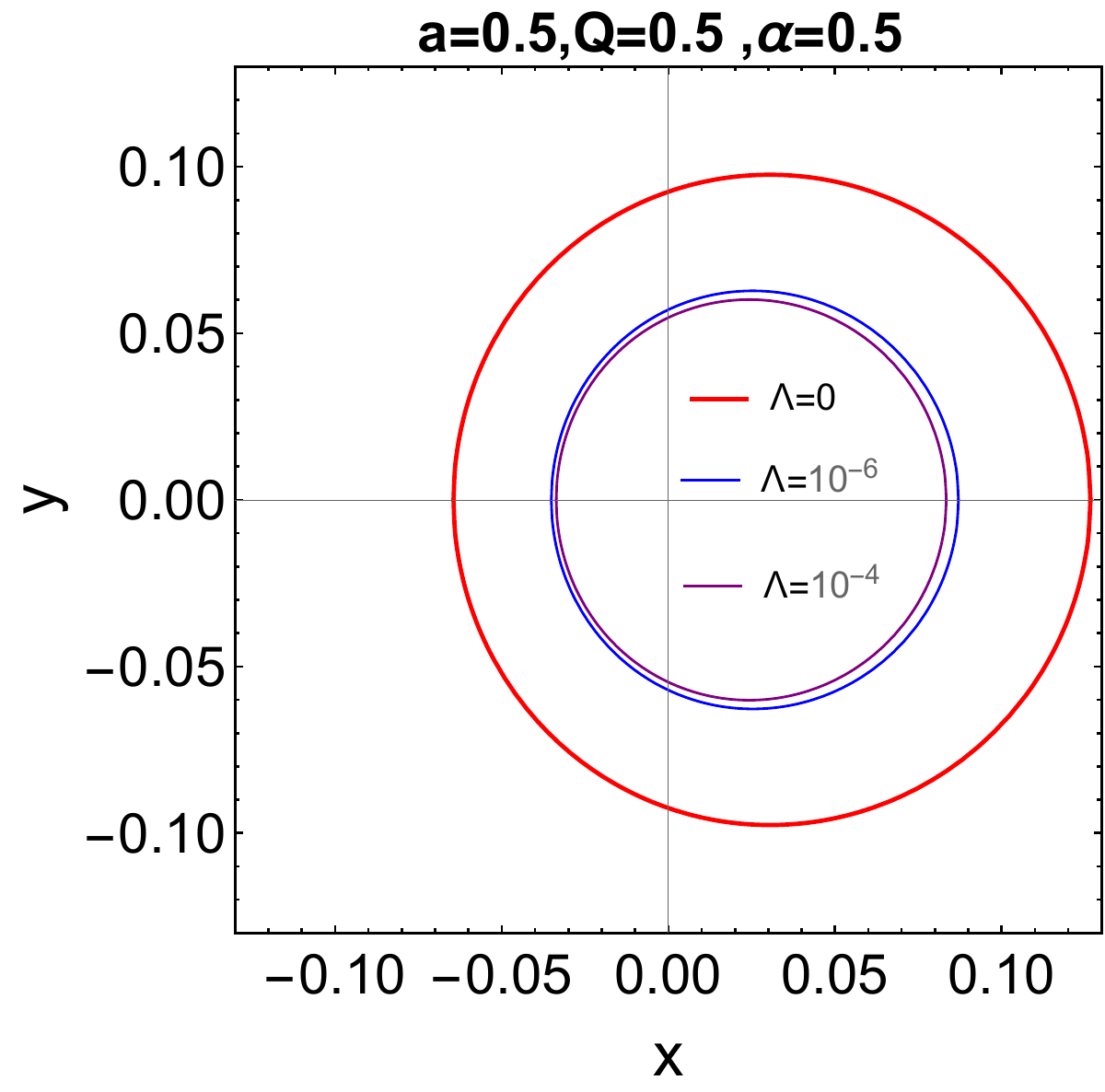}\\
\caption{Boundary of the black hole shadow for different values of the $\Lambda$, setting $a=0.5, Q=0.5, \alpha=0.5$ and $M=1$.}
\label{fig10}
\end{minipage}\\
\begin{minipage}[t]{0.48\linewidth}
\centering
\includegraphics[width=3in]{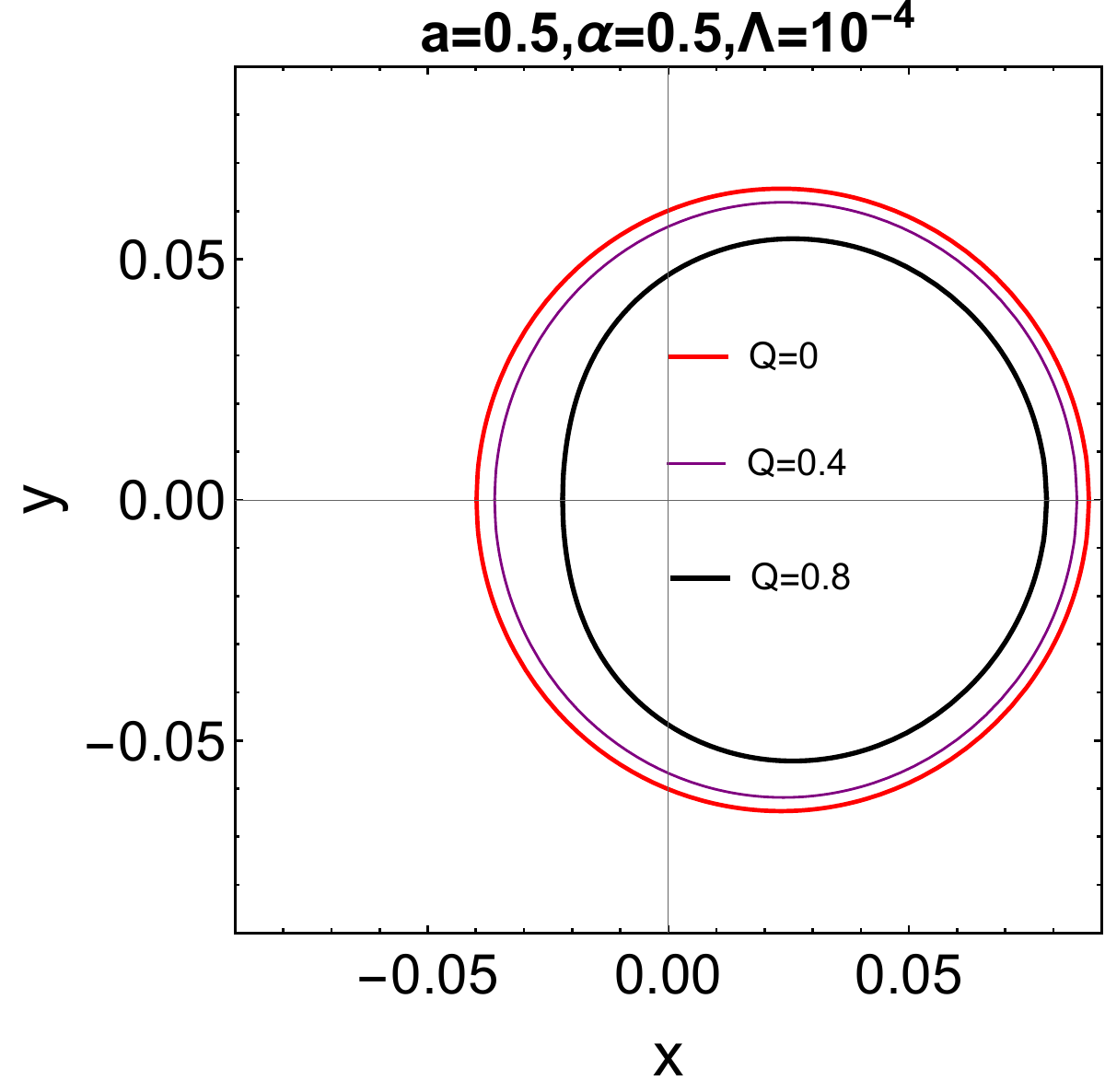}
\caption{Boundary of the black hole shadow for different values of the $Q$, setting $a=0.5, \alpha=0.5, \Lambda=10^{-4}$ and $M=1$.}
\label{fig11}
\end{minipage}
\hspace{0.1cm}
\begin{minipage}[t]{0.48\linewidth}
\centering
\includegraphics[width=3.2in]{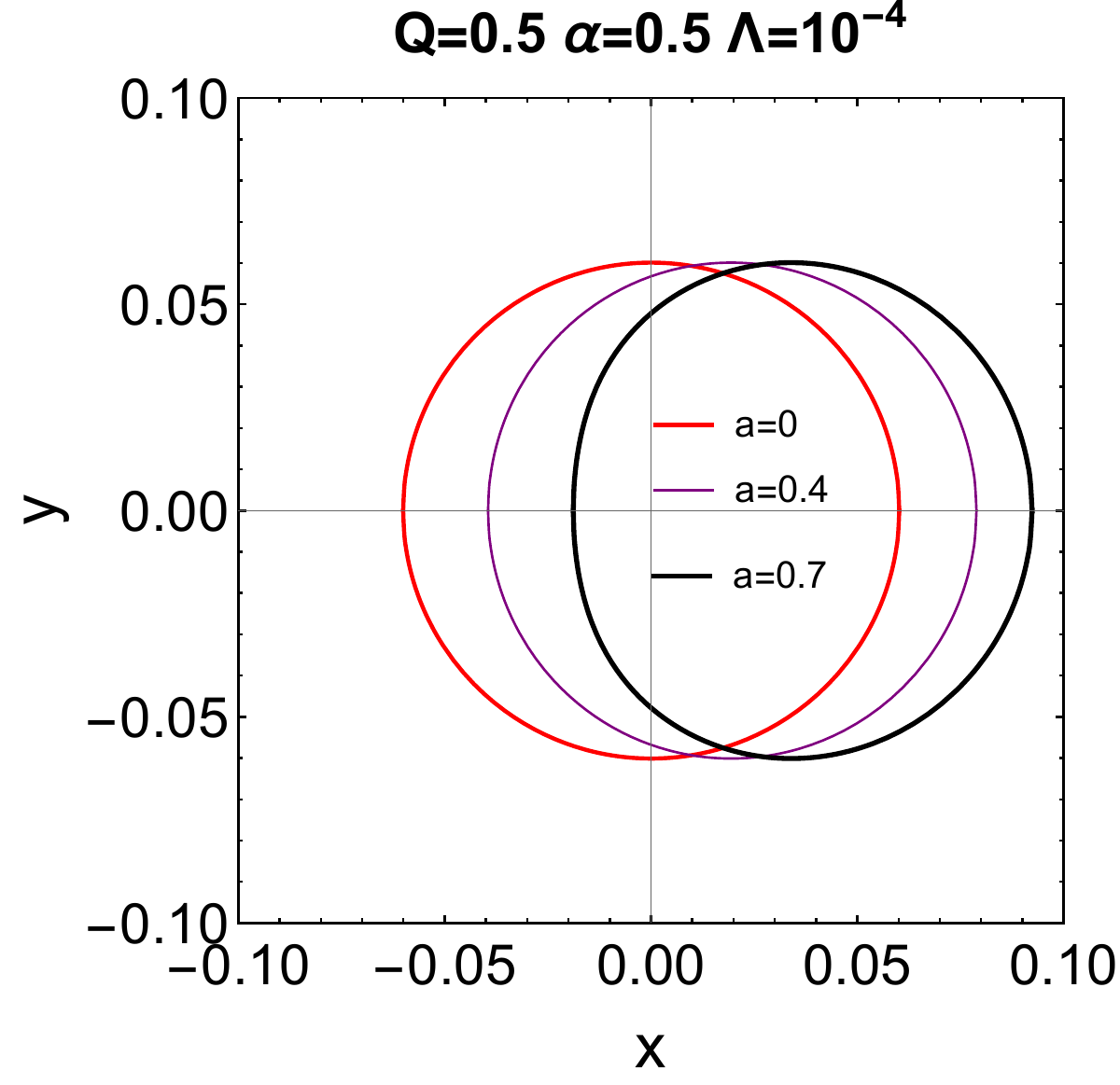}\\
\caption{Boundary of the black hole shadow for different values of the $a$, setting $Q=0.5, \alpha=0.5, \Lambda=10^{-4}$ and $M=1$.}
\label{fig12}
\end{minipage}
\end{figure}
Following the reference \cite{Grenzebach:2014fha}, transforming from the celestial coordinates $(\alpha(r_s),\beta(r_s))$ to the standard Cartesian coordinates $(x(r_s),y(r_s))$, which are given by
\begin{align}
x\left(r_{s}\right)&=-2 \tan \left(\frac{\alpha\left(r_{s}\right)}{2}\right) \sin \left(\beta\left(r_{s}\right)\right),\label{xrs} \\
y\left(r_{s}\right)&=-2 \tan \left(\frac{\alpha\left(r_{s}\right)}{2}\right) \cos \left(\beta\left(r_{s}\right)\right).\label{yrs}
\end{align}

We choose that the observer is located at \( r_O=50 \) and \( \theta=\pi/2 \). The boundaries of the new rotating BHs are plotted for the different values of \( \alpha \), \( \Lambda \), \( Q \), and \( a \) in Figs. \ref{fig9}-\ref{fig12}, respectively. It is obvious that the boundaries of the rotating BHs nonlinearly change in Fig. \ref{fig9} when \( \alpha \) increases. The boundaries of the rotating BHs decrease with the increase of \( \Lambda \) and \( Q \) in Figs. \ref{fig10}-\ref{fig11}. From Fig. \ref{fig12}, the shape of the BH shadows deviates from the circle as \( a \) increases.

\subsection{The deformation of the BH shadow}

To describe the distortion and size of the charged rotating BH with a cosmological constant immersed in PFDM Eq.(\ref{knd8}), we first study two characteristic observables, $R_s$ and $\delta_s$, which were proposed by Hioki and Maeda \cite{Hioki:2009na}. Here, $R_s$ is the radius of the reference circle for the distorted shadow, and $\delta_s$ is the deviation of the left edge of the shadow from the reference circle boundary. For convenience, we denote the top, bottom, right, and left of the reference circle as $(x_t,y_t)$, $(x_b,y_b)$, $(x_r,0)$, and $(\tilde{x}_l,0)$, respectively, and $(x_l,0)$ is the leftmost edge of the shadow []. Subsequently, the definitions of the characteristic observables are \cite{Hioki:2009na}
\begin{align}
R_{s}=&\frac{\left(x_{t}-x_{r}\right)^{2}+y_{t}^{2}}{2\left(x_{r}-x_{t}\right)},\label{Rs}\\
\delta_{s}=&\frac{\left|x_{l}-\tilde{x}_{l}\right|}{R_s} .\label{delta}
\end{align}

\begin{figure}[htbp]
\centering
\includegraphics[scale=0.36]
{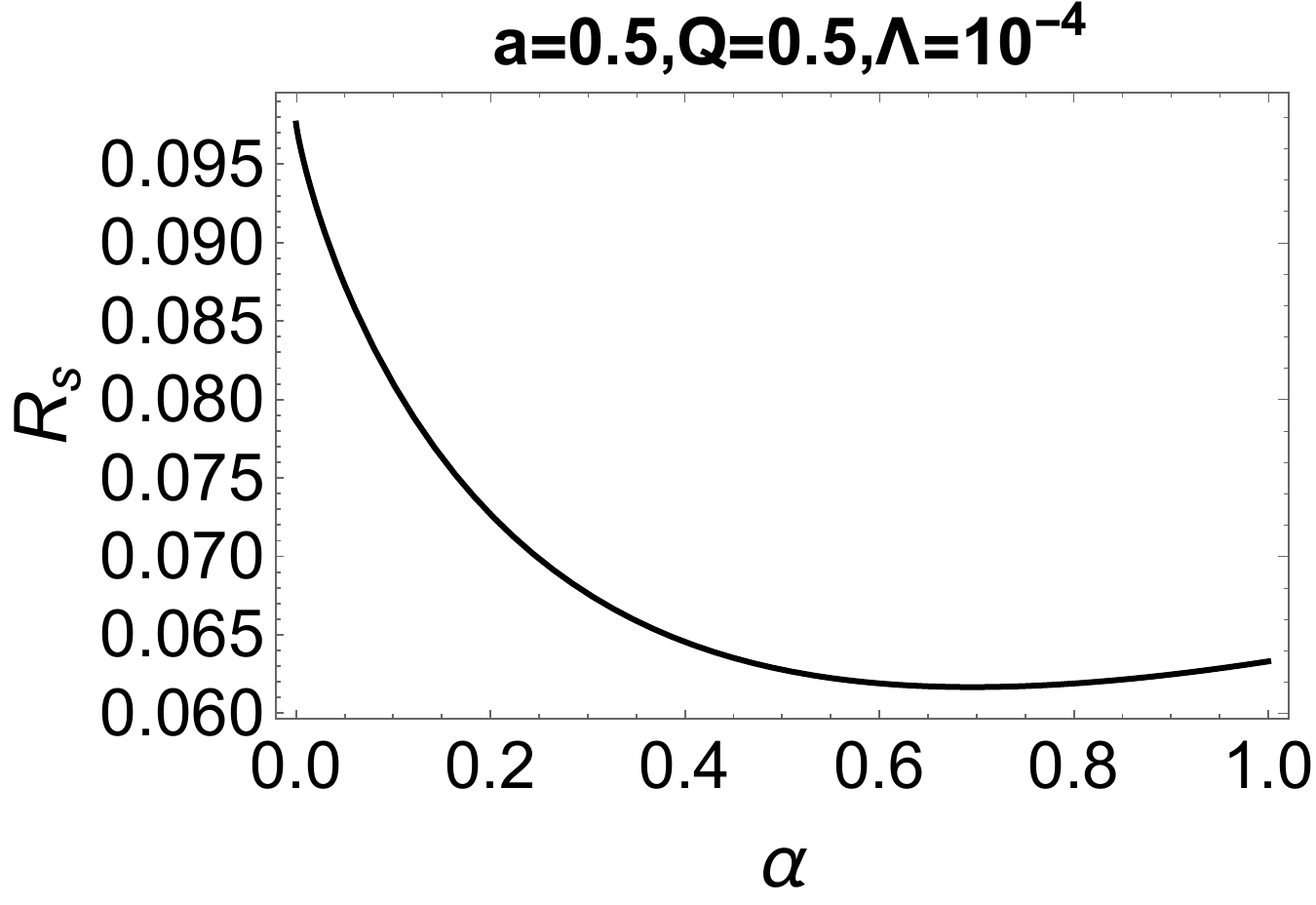}
\includegraphics[scale=0.35]{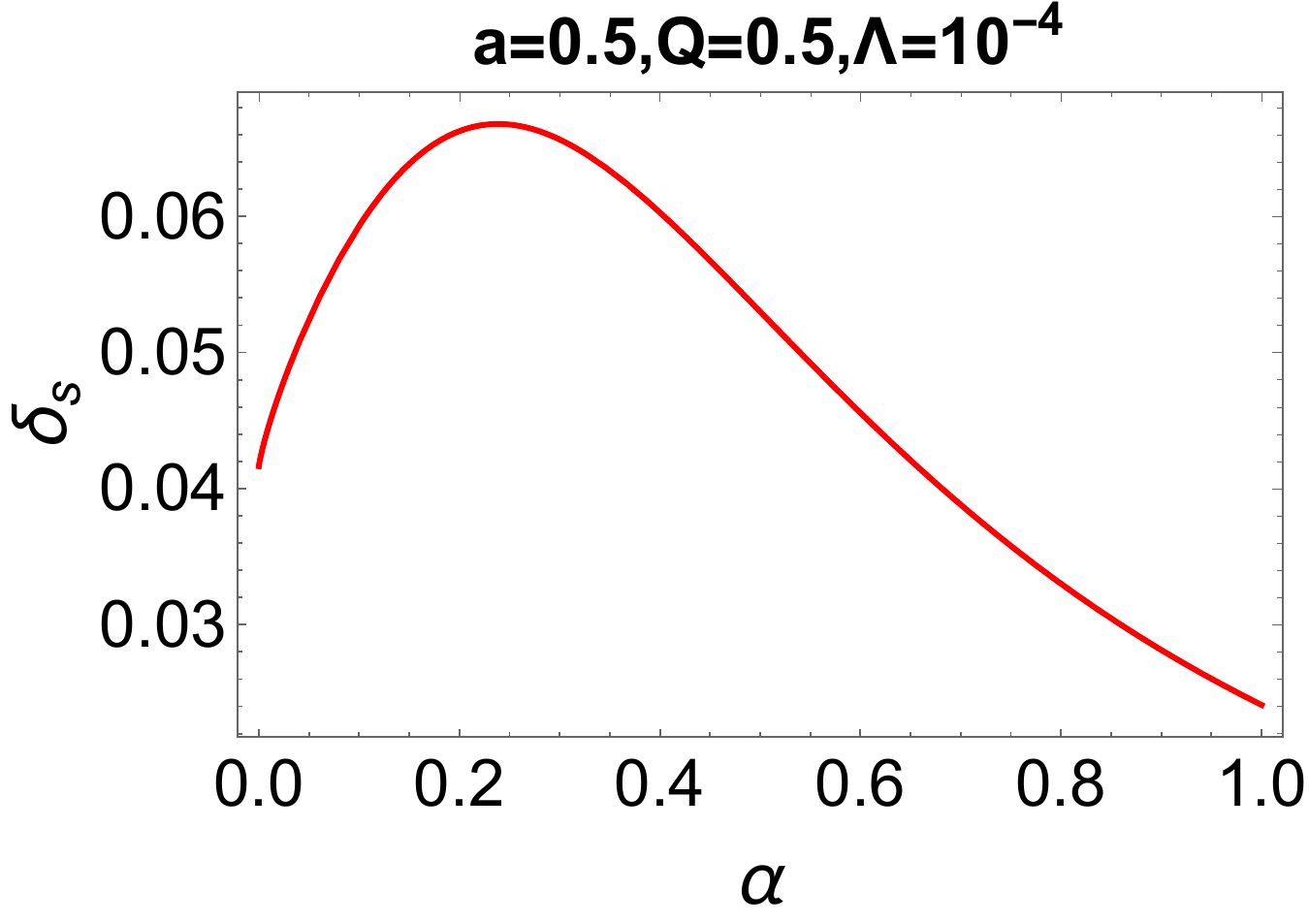}
\caption{The left-hand is the $R_s$ as the function of the $\alpha$, while the right-hand is the $\delta_s$ as the function of the $\alpha$, setting $a=0.5, Q=0.5, \Lambda=10^{-4}$ and $M=1$.}
\label{fig13}
\end{figure}
\begin{figure}[htbp]
\centering
\includegraphics[scale=0.36]
{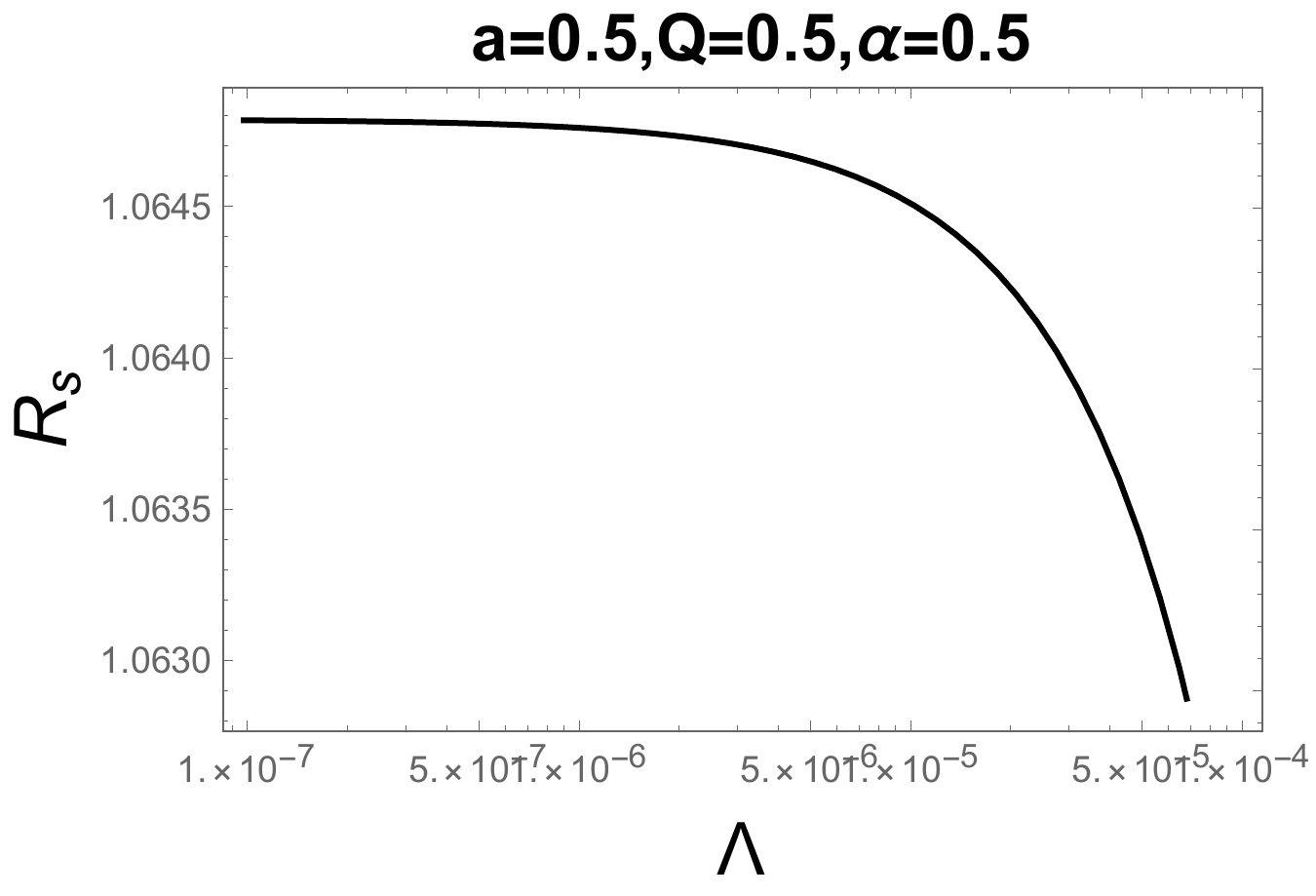}
\includegraphics[scale=0.35]{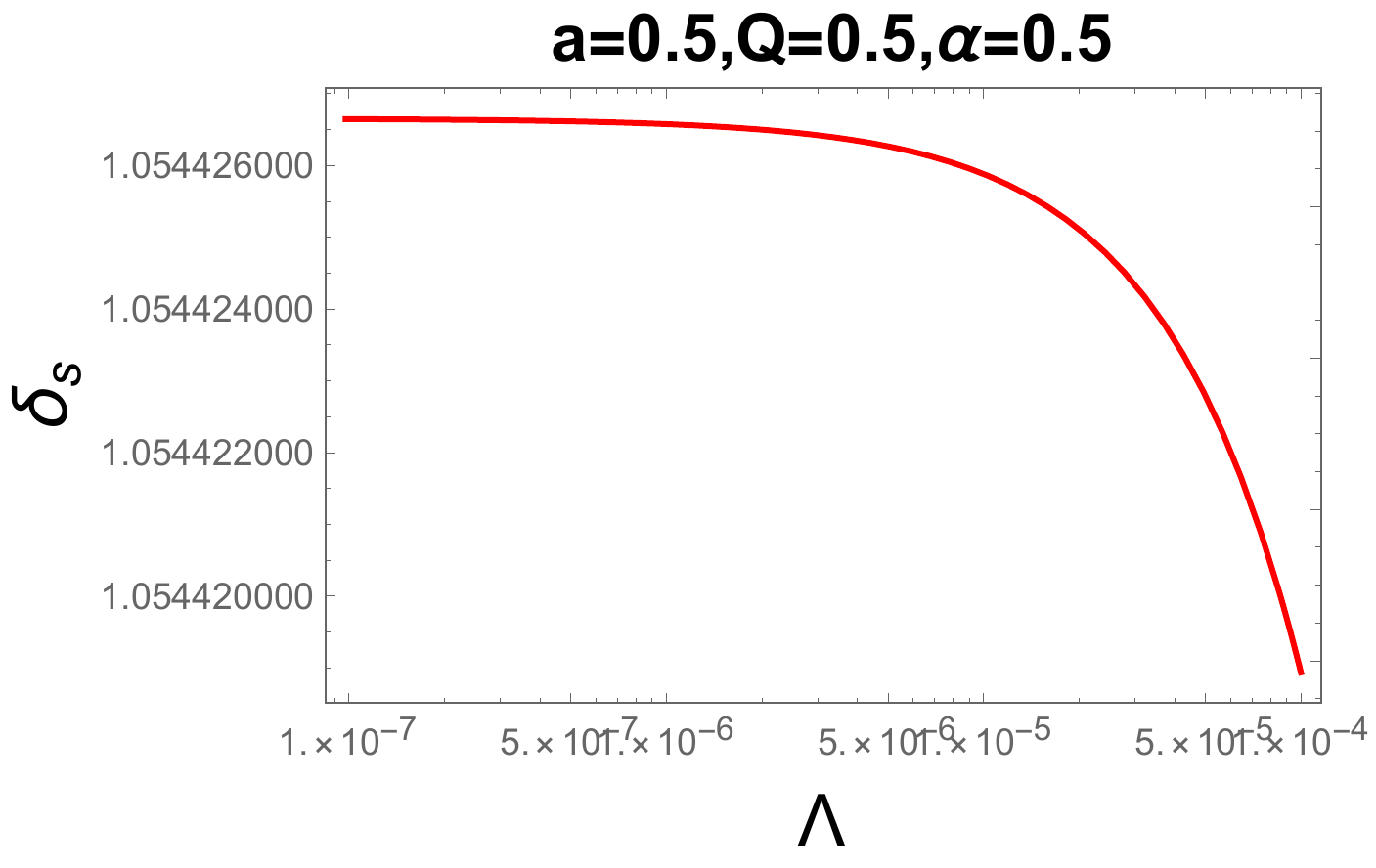}
\caption{The left-hand denotes the $R_s$ as the function of the $\Lambda$, while the right-hand denotes the $\delta_s$ as the function of the $\Lambda$, setting $a=0.5, Q=0.5, \alpha=0.5$ and $M=1$.}
\label{fig14}
\end{figure}
\begin{figure}[htbp]
\centering
\includegraphics[scale=0.36]
{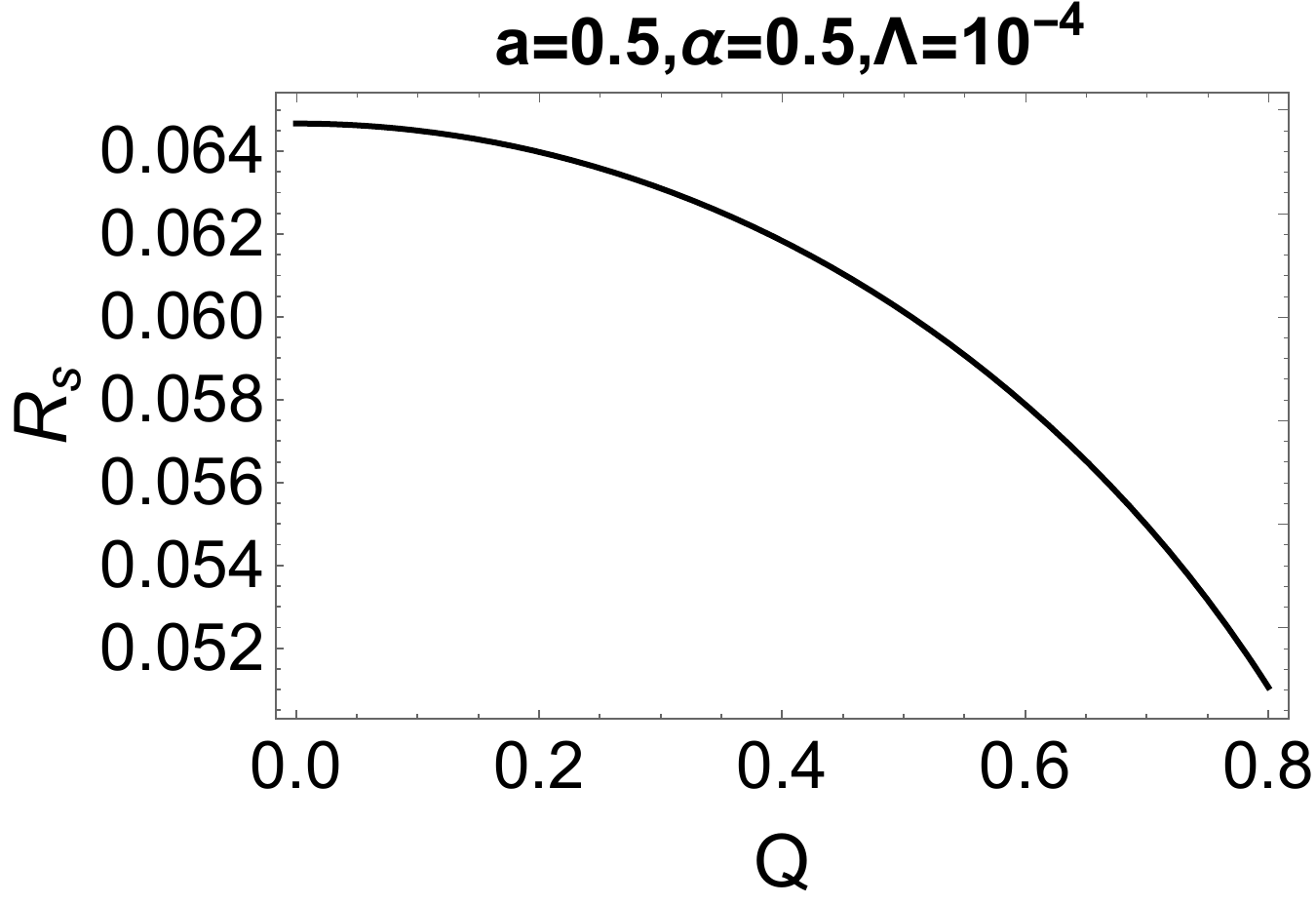}
\includegraphics[scale=0.35]{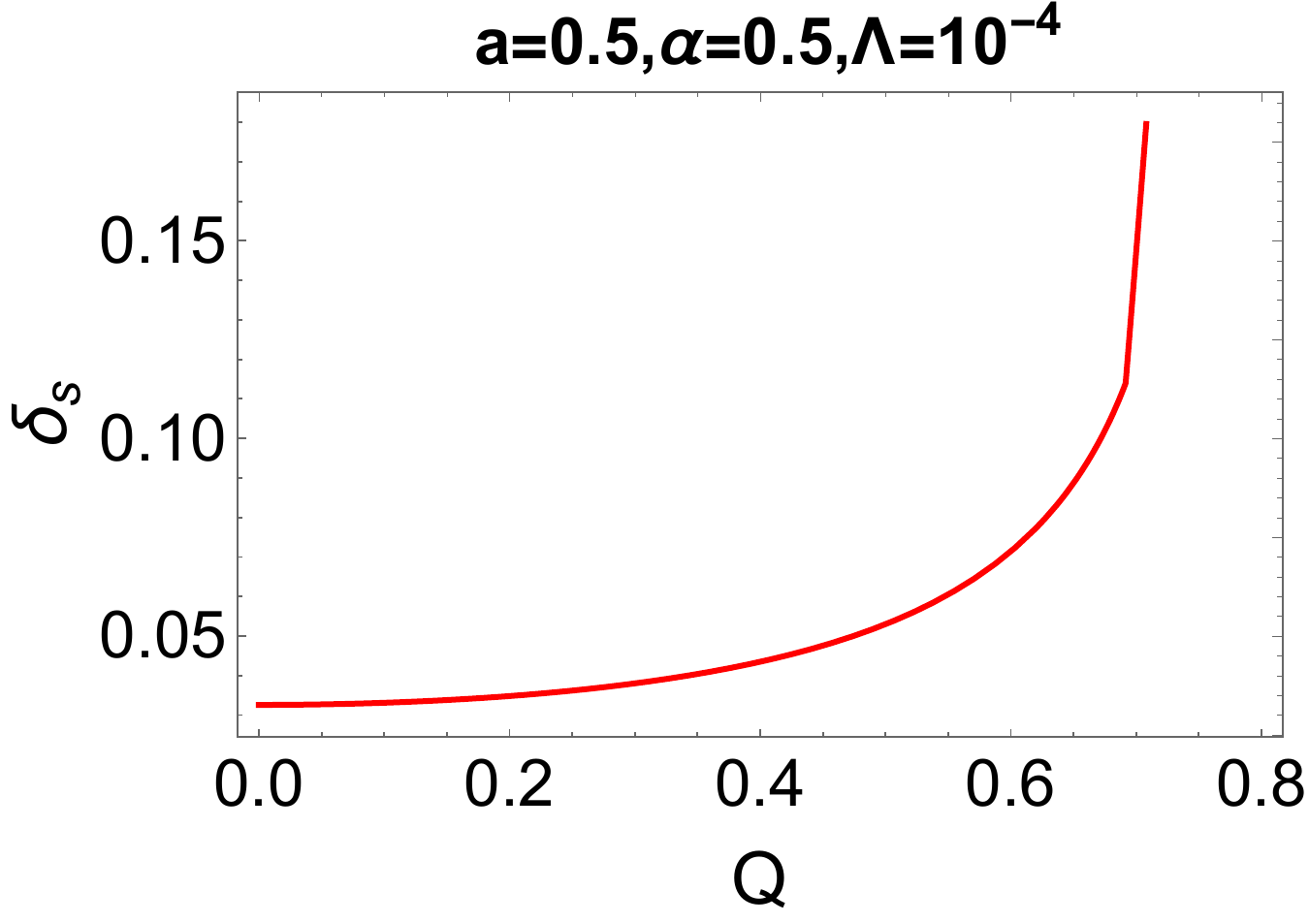}
\caption{The left-hand represents the $R_s$ as the function of the $Q$, while the right-hand represents the $\delta_s$ as the function of the $Q$, setting $a=0.5, \alpha=0.5, \Lambda=10^{-4}$ and $M=1$.}
\label{fig15}
\end{figure}
\begin{figure}[htbp]
\centering
\includegraphics[scale=0.36]
{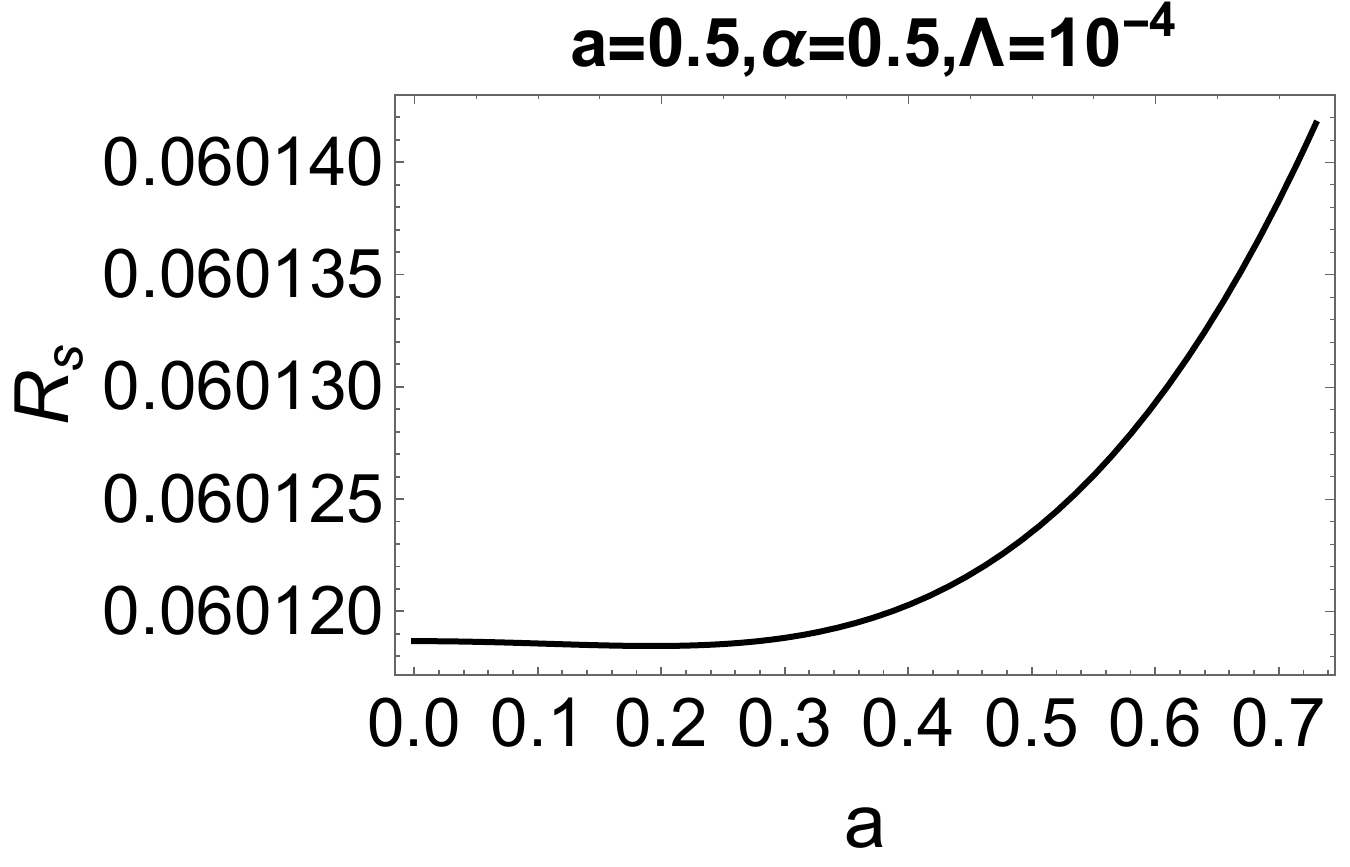}
\includegraphics[scale=0.35]{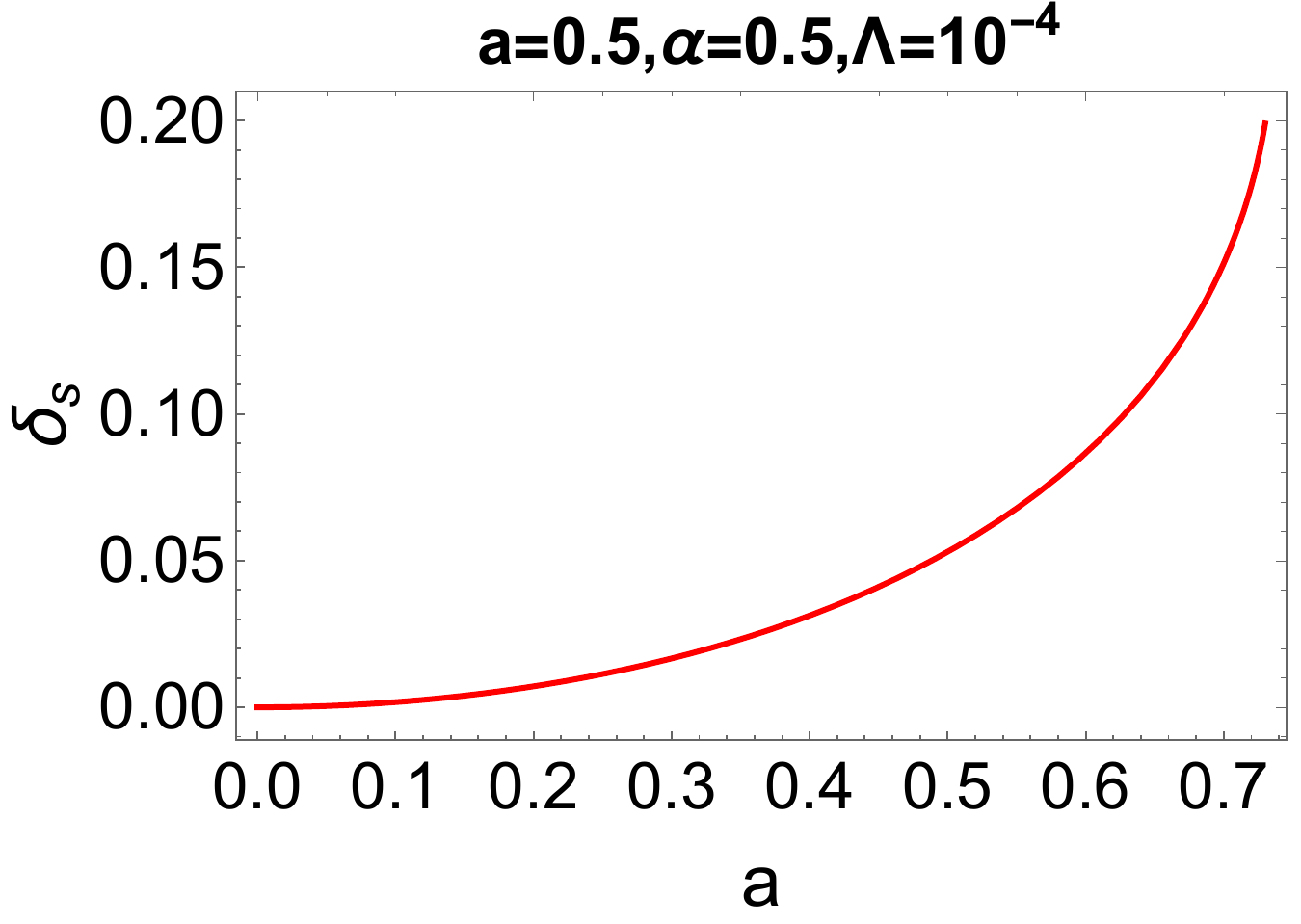}
\caption{The left-hand represents the $R_s$ as the function of the $a$, while the right-hand represents the $\delta_s$ as the function of the $a$, setting $Q=0.5, \alpha=0.5, \Lambda=10^{-4}$ and $M=1$.}
\label{fig16}
\end{figure}
We have plotted \( R_s \) and \( \delta_s \) as functions of the four parameters \( \alpha, \Lambda, Q \), and \( a \) in Figs. \ref{fig13}-\ref{fig16}, respectively. When \( \alpha \) gradually increases, \( R_s \) continuously decreases, while \( \delta_s \) first increases and then decreases in Fig. \ref{fig13}. \( R_s \) and \( \delta_s \) decrease with the increase of \( \Lambda \) in Fig. \ref{fig14}. It clearly shows that \( R_s \) decreases as \( Q \) increases in Fig. \ref{fig15}, but \( \delta_s \) gradually increases. From Fig. \ref{fig16}, we can see that \( R_s \) and \( \delta_s \) increase with the increase of \( a \).

\section{Conclusion and discussion}
\label{section4}

This study explored the shadow of a rotating charged BH within the context of PFDM and a cosmological constant. Using the gravitational decoupling method, we derived a solution for a charged, spherically symmetric BH in PFDM and extended it to the rotating case. Our analysis highlighted the substantial effects of the PFDM parameter \(\alpha\) on the event horizons, effective potential functions, and deformations of the BH shadows, surpassing the influences of the cosmological constant \(\Lambda\), electric charge \(Q\), and rotation parameter \(a\).

The study demonstrated that increasing \(\alpha\) nonlinearly alters the boundaries of the rotating BH shadows, while \(\Lambda\) and \(Q\) tend to decrease these boundaries. The rotation parameter \(a\) causes the BH shadows to deviate from circularity. These findings suggest that PFDM plays a pivotal role in influencing the observable features of BHs, offering a new perspective on how different parameters affect BH characteristics.

The numerical analysis provides a comprehensive understanding of how these parameters shape the BH shadows, offering valuable insights into the interplay between gravity, dark matter, and dark energy. This work is expected to inform future efforts to align theoretical models with observations of celestial bodies in dark matter and energy environments. In summary, the shadow of a rotating charged BH in PFDM, especially with a cosmological constant, remains a compelling topic for further research. Our results not only enhance theoretical knowledge of BH physics but also promise to guide future observational studies probing the nature of dark matter and dark energy in BH environments.

\begin{acknowledgments}
	This work is supported in part by NSFC Grant No. 12165005.
\end{acknowledgments}


\end{document}